\documentclass[letterpaper, 10pt, conference]{ieeeconf}      
\pdfoutput=1

\IEEEoverridecommandlockouts                              

\usepackage[main=american,ngerman]{babel}
\addto\captionsenglish{}
\addto\captionsenglish{}
\usepackage{amsmath,amssymb,amsfonts}
\usepackage{algorithmic}
\usepackage{todonotes}
\usepackage{tikzsymbols}
\usepackage{blindtext}
\usepackage{csquotes}
\usepackage{dblfloatfix}
\usepackage{graphicx}
\usepackage{framed}
\usepackage{textcomp}
\usetikzlibrary{shadows.blur,arrows.meta}
\usepackage{xcolor}
\usepackage{siunitx}
\usepackage{orcidlink}
\usepackage{nicematrix}
\usepackage{arydshln}

\usepackage[strict]{changepage}

\usepackage{framed}
\definecolor{grauquote}{RGB}{250,250,250}

\usepackage{cleveref}
\crefname{figure}{Fig.}{Fig.}
\crefname{equation}{}{}
\Crefname{equation}{Equation}{Equations}

\newcommand{\unicaragil}{UNICAR\emph{agil}}
\newcommand{\autotech}{AUTOtech.\emph{agil}}
\newcommand{\autotaxi}{\emph{auto}TAXI}
\newcommand{\autoelf}{\emph{auto}ELF}
\newcommand{\autocargo}{\emph{auto}CARGO}
\newcommand{\autoshuttle}{\emph{auto}SHUTTLE}

\definecolor{rotLogo}{RGB}{128,29,26}
\definecolor{graux}{RGB}{240,240,240}
\definecolor{tableGreen}{RGB}{222,236,224}
\definecolor{greenAct}{RGB}{231,241,232}
\definecolor{blueCite}{RGB}{155, 176, 193}
\definecolor{beigeRef}{RGB}{234, 223, 180}
\definecolor{blaux}{RGB}{52,98,108}

\definecolor{blueAct}{RGB}{157,208,225}
\definecolor{orangeAct}{RGB}{229,126,0}
\definecolor{greyAct}{RGB}{207,207,207}

\hypersetup{hidelinks,colorlinks,linkcolor=blaux,citecolor=blaux,urlcolor=blaux}

\newcommand{\GreyBox}{\raisebox{-1.1mm}{\tikz{\filldraw[blur shadow={shadow blur steps=10},shadow scale=.95,rounded corners = 2pt,greyAct,solid,line width=.5pt,draw=white](0,0) rectangle ++ (20mm,3.4mm) node[pos=.5,black] {\textbf{Grey boxes}};}}\hspace*{0.03cm}}
\newcommand{\OrangeBox}{\raisebox{-1.1mm}{\tikz{\filldraw[blur shadow={shadow blur steps=10},shadow scale=.95,orangeAct,solid,line width=.5pt,draw=white](0,0) rectangle ++ (20mm,3.4mm) node[pos=.5,black] {\textbf{Orange boxes}};}}\hspace*{0.03cm}}
\newcommand{\BlueBox}{\raisebox{-1.1mm}{\tikz{\filldraw[blur shadow={shadow blur steps=10},shadow scale=.95,blueAct,solid,line width=.5pt,draw=white](0,0) rectangle ++ (20mm,3.4mm) node[pos=.5,black] {\textbf{Blue boxes}};}}\hspace*{0.03cm}}

\usepackage[normalem]{ulem}
\newcommand\reduline{\bgroup\markoverwith{\textcolor{rotLogo}{\rule[-0.8ex]{2pt}{.5pt}}}\ULon}
\newcommand{\ReqBox}{\raisebox{-0.7mm}{\includegraphics[scale=0.08]{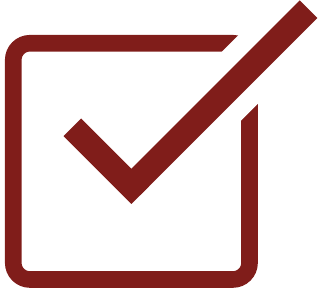}}}
\newcommand{\RequirementLine}[1]{\ReqBox~\hspace{0em}\reduline{\textbf{#1}\hfill}\\[.2em]\noindent}

\newcommand{\SecLine}[1]{{\vspace{.6pt}\noindent\hspace{0em}\reduline{\textit{#1}\hfill}}\\[.3em]\noindent}

\newcommand{\GeneralizationRelShip}{\tikz{\draw[->,arrows={-Triangle[open,scale=1.2]},line width=.7pt](0,0) -- ++ (.6,0);}}
\newcommand{\AggregationRelShip}{\raisebox{0.35mm}{\tikz{\draw[->,arrows={-Diamond[open,scale=1.2]},line width=.7pt](0,0) -- ++ (.6,0);}}}
\newcommand{\Actor}{\raisebox{-0.95mm}{\includegraphics[scale=0.45]{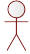}}}

\let\oldmaketitle\maketitle
\renewcommand{\maketitle}{\oldmaketitle\setcounter{footnote}{1}}

\usepackage[style=ieee,
backend=biber,
minnames=1,
maxcitenames=2,
maxbibnames=5,
doi=true,
isbn=false,
url=false,
natbib=true,
clearlang=true,
date=year
]{biblatex}
\usepackage{xpatch}
\usepackage{xstring}

\DeclareSourcemap{
	\maps[datatype=bibtex]{
		\map{
			\step[fieldsource=langid, match=\regexp{\A(n)?((swiss)?german|austrian)\Z}, final]
			\step[fieldset=language, fieldvalue={(in German)}]
		}
		\map{
			\step[fieldsource=langid, match=pinyin, final]
			\step[fieldset=language, fieldvalue={(in Chinese)}]
		}
		\map{
			\step[fieldsource=langid, match=japanese, final]
			\step[fieldset=language, fieldvalue={(in Japanese)}]
		}
		\map{
			\step[fieldsource=langid, match=ko, final]
			\step[fieldset=language, fieldvalue={(in Korean)}]
		}
	}
}

\xpatchbibdriver{inproceedings}					 	
{\usebibmacro{publisher+location+date}}			
{\IfSubStr{\strfield{booktitle}}{\strfield{year}}{\usebibmacro{publisher+location}}{\usebibmacro{publisher+location+date}}} 
{} 
{\typeout{There was an error patching biblatex-ieee (specifically, ieee.bbx's @inproceedings driver)}} 

\newbibmacro*{publisher+location}{%
	\printlist{location}%
	\iflistundef{publisher}
	{\setunit*{\addcomma\space}}
	{\setunit*{\addcolon\space}}%
	\printlist{publisher}%
	\newunit}

\xpatchbibdriver{online}
{\printtext[parens]{\usebibmacro{date}}}
{\iffieldundef{year}
	{}
	{\printtext[parens]{\usebibmacro{date}}}}
{}
{\typeout{There was an error patching biblatex-ieee (specifically, ieee.bbx's @online driver)}}

\DeclareSourcemap{
	\maps[datatype=biber]{
		\map{
			\step[fieldsource=note, final]
			\step[fieldset=addendum, origfieldval, final]
			\step[fieldset=note, null]
		}
	}
}

\DeclareBibliographyDriver{standard}{%
	\usebibmacro{bibindex}%
	\usebibmacro{begentry}%
	\usebibmacro{maintitle+title}%
	\setunit{\addcomma\addspace}%
	\usebibmacro{publisher+type+number}%
	\setunit{\labelnamepunct}\newblock
	\IfSubStr{\strfield{number}}{\strfield{year}}{}{\usebibmacro{date}} 
	\newunit\newblock
	\usebibmacro{finentry}%
}

\newbibmacro*{publisher+type+number}{%
	\printtext{%
		\printlist{publisher}
		\iflistundef{publisher}
		{\space}{}%
		\iffieldundef{type}
		{Standard}
		{\printfield{type}}
		\printfield{number}
	}%
}

\xpatchbibdriver{report}					 	
{\usebibmacro{date}}			
{\IfSubStr{\strfield{number}}{\strfield{year}}{}{\usebibmacro{date}}} 
{} 
{\typeout{There was an error patching biblatex-ieee (specifically, ieee.bbx's @report driver)}} 

\DeclareBibliographyDriver{software}{%
	\usebibmacro{bibindex}%
	\usebibmacro{begentry}%
	\usebibmacro{maintitle+title}%
	\setunit{\addcomma\addspace}%
	\usebibmacro{version+year}%
	\setunit{\labelnamepunct}\newblock
	\usebibmacro{author}%
	\setunit{\addcomma\addspace}%
	\newunit\newblock
	\usebibmacro{url+urldate}%
	\newunit\newblock
	\usebibmacro{finentry}%
}

\newbibmacro*{version+year}{%
	\printtext{%
		\iffieldundef{version}{\iffieldundef{year}{}{\printfield{year}}}{\printfield{version}}
		\printfield{number}
		\printlist{publisher}
		\iflistundef{publisher}{\space}{}%
	}%
}

\DeclareSourcemap{
	\maps[datatype=bibtex]{
		\map[overwrite, foreach={booktitle,journaltitle,eventtitle,series,publisher,institution,type}]{
			\step[fieldsource=\regexp{$MAPLOOP}, match={XXX}, replace={Acad. Serbe Sci. Arts Glas Cl. Sci. Tech.}]
			\step[fieldsource=\regexp{$MAPLOOP}, match={XXX}, replace={Acta Acust.}]
			\step[fieldsource=\regexp{$MAPLOOP}, match={XXX}, replace={Acta Astron. (Poland)}]
			\step[fieldsource=\regexp{$MAPLOOP}, match={XXX}, replace={Acta Astron. Sin. (China)}]
			\step[fieldsource=\regexp{$MAPLOOP}, match={XXX}, replace={Acta Astronaut. (U.K.)}]
			\step[fieldsource=\regexp{$MAPLOOP}, match={XXX}, replace={Acta Astrophys. Sin. (China)}]
			\step[fieldsource=\regexp{$MAPLOOP}, match={XXX}, replace={Acta Autom. Sin.}]
			\step[fieldsource=\regexp{$MAPLOOP}, match={XXX}, replace={Acta Cienc. Indica Math.}]
			\step[fieldsource=\regexp{$MAPLOOP}, match={XXX}, replace={Acta Cienc. Indica Phys.}]
			\step[fieldsource=\regexp{$MAPLOOP}, match={XXX}, replace={Acta Crystallogr. A, Found. Crystallogr.}]
			\step[fieldsource=\regexp{$MAPLOOP}, match={XXX}, replace={Acta Crystallogr. B, Struct. Sci.}]
			\step[fieldsource=\regexp{$MAPLOOP}, match={XXX}, replace={Acta Crystallogr. C, Cryst. Struct. Commun.}]
			\step[fieldsource=\regexp{$MAPLOOP}, match={XXX}, replace={Acta Electron. (France)}]
			\step[fieldsource=\regexp{$MAPLOOP}, match={XXX}, replace={Acta Cyberno}]
			\step[fieldsource=\regexp{$MAPLOOP}, match={XXX}, replace={Acta Electron. Sin. (China)}]
			\step[fieldsource=\regexp{$MAPLOOP}, match={XXX}, replace={Acta Geod. Geophys. Montan. Hung.}]
			\step[fieldsource=\regexp{$MAPLOOP}, match={XXX}, replace={Acta Geophys. Pol.}]
			\step[fieldsource=\regexp{$MAPLOOP}, match={XXX}, replace={Acta Geophys. Sin. (China)}]
			\step[fieldsource=\regexp{$MAPLOOP}, match={XXX}, replace={Acta Geophys. Sin. (USA)}]
			\step[fieldsource=\regexp{$MAPLOOP}, match={XXX}, replace={Acta Metall.}]
			\step[fieldsource=\regexp{$MAPLOOP}, match={XXX}, replace={Acta Mex. Cienc. Tecnol.}]
			\step[fieldsource=\regexp{$MAPLOOP}, match={XXX}, replace={Acta Phys. Hung.}]
			\step[fieldsource=\regexp{$MAPLOOP}, match={XXX}, replace={Acta Phys. Pol. A}]
			\step[fieldsource=\regexp{$MAPLOOP}, match={XXX}, replace={Acta Phys. Pol. B}]
			\step[fieldsource=\regexp{$MAPLOOP}, match={XXX}, replace={Acta Phys. Sin.}]
			\step[fieldsource=\regexp{$MAPLOOP}, match={XXX}, replace={Acta Phys. Slovaca}]
			\step[fieldsource=\regexp{$MAPLOOP}, match={XXX}, replace={Acta Politec. Mex.}]
			\step[fieldsource=\regexp{$MAPLOOP}, match={XXX}, replace={Acta Polytech. Scand. Appl. Phys. Ser.}]
			\step[fieldsource=\regexp{$MAPLOOP}, match={XXX}, replace={Acta Polytech. Scand. Chem. Technol.}]
			\step[fieldsource=\regexp{$MAPLOOP}, match={XXX}, replace={Metall. Ser.}]
			\step[fieldsource=\regexp{$MAPLOOP}, match={XXX}, replace={Acta Polytech. Scand. Electr. Eng. Ser.}]
			\step[fieldsource=\regexp{$MAPLOOP}, match={XXX}, replace={Acta Polytech. Scand. Math. Comput. Sci. Ser.}]
			\step[fieldsource=\regexp{$MAPLOOP}, match={XXX}, replace={Acta Polytech. Scand. Mech. Eng. Ser.}]
			\step[fieldsource=\regexp{$MAPLOOP}, match={XXX}, replace={Acta Seismol. Sin.}]
			\step[fieldsource=\regexp{$MAPLOOP}, match={XXX}, replace={Acta Tech. Acad. Sci. Hung.}]
			\step[fieldsource=\regexp{$MAPLOOP}, match={XXX}, replace={Acta Tech. CSAV}]
			\step[fieldsource=\regexp{$MAPLOOP}, match={XXX}, replace={Acustica}]
			\step[fieldsource=\regexp{$MAPLOOP}, match={XXX}, replace={(AEU) Arch. Elektr. Ubertragung}]
			\step[fieldsource=\regexp{$MAPLOOP}, match={XXX}, replace={Akust. Zh.}]
			\step[fieldsource=\regexp{$MAPLOOP}, match={XXX}, replace={Algorithmica}]
			\step[fieldsource=\regexp{$MAPLOOP}, match={XXX}, replace={Alta Freq.}]
			\step[fieldsource=\regexp{$MAPLOOP}, match={XXX}, replace={An. Acad. Bras. Cienc.}]
			\step[fieldsource=\regexp{$MAPLOOP}, match={XXX}, replace={An. Fis.}]
			\step[fieldsource=\regexp{$MAPLOOP}, match={XXX}, replace={An. Mec. Electr.}]
			\step[fieldsource=\regexp{$MAPLOOP}, match={XXX}, replace={Angew. Inform.}]
			\step[fieldsource=\regexp{$MAPLOOP}, match={XXX}, replace={Ann. Inst. Henri Poincare Phys. Theor.}]
			\step[fieldsource=\regexp{$MAPLOOP}, match={XXX}, replace={Ann. Soc. Sci. Brux. I, Sci. Math. Astron. Phys.}]
			\step[fieldsource=\regexp{$MAPLOOP}, match={XXX}, replace={Arch. Elektr. Uebertrag. (AEU)}] 
			\step[fieldsource=\regexp{$MAPLOOP}, match={XXX}, replace={Arch. Elektron. Uebertrag. Tech.}] 
			\step[fieldsource=\regexp{$MAPLOOP}, match={XXX}, replace={Arch. Elektrotech. (Poland)}]
			\step[fieldsource=\regexp{$MAPLOOP}, match={XXX}, replace={Arch. Elektrotech. (Germany)}]
			\step[fieldsource=\regexp{$MAPLOOP}, match={XXX}, replace={Ark. Fys. Semin. Trondheim}]
			\step[fieldsource=\regexp{$MAPLOOP}, match={XXX}, replace={Astrofizika}]
			\step[fieldsource=\regexp{$MAPLOOP}, match={XXX}, replace={Astron. Nachr. (Germany)}]
			\step[fieldsource=\regexp{$MAPLOOP}, match={XXX}, replace={Astron. Tidsskr.}]
			\step[fieldsource=\regexp{$MAPLOOP}, match={XXX}, replace={Astron. Vestn.}]
			\step[fieldsource=\regexp{$MAPLOOP}, match={XXX}, replace={Astron. Zh.}]
			\step[fieldsource=\regexp{$MAPLOOP}, match={XXX}, replace={Atomwirtsch.-Atomtech.}]
			\step[fieldsource=\regexp{$MAPLOOP}, match={XXX}, replace={Atti Accad. Sci. Ist. Bologna CI. Sci. Fis.}]
			\step[fieldsource=\regexp{$MAPLOOP}, match={XXX}, replace={Rend. XIII}]
			\step[fieldsource=\regexp{$MAPLOOP}, match={XXX}, replace={Atti Accad. Sci. Torino I, CI. Sci. Fis. Math. Nat.}]
			\step[fieldsource=\regexp{$MAPLOOP}, match={XXX}, replace={Autom. Strum.}]
			\step[fieldsource=\regexp{$MAPLOOP}, match={XXX}, replace={Autom. Tech. Prax.}]
			\step[fieldsource=\regexp{$MAPLOOP}, match={XXX}, replace={Automatica}]
			\step[fieldsource=\regexp{$MAPLOOP}, match={XXX}, replace={Automatie}]
			\step[fieldsource=\regexp{$MAPLOOP}, match={XXX}, replace={Automatika}]
			\step[fieldsource=\regexp{$MAPLOOP}, match={XXX}, replace={Automatisierungstechnik}]
			\step[fieldsource=\regexp{$MAPLOOP}, match={XXX}, replace={Automatizace}]
			\step[fieldsource=\regexp{$MAPLOOP}, match={XXX}, replace={Automedica}]
			\step[fieldsource=\regexp{$MAPLOOP}, match={XXX}, replace={Avtom. Telemekh.}]
			\step[fieldsource=\regexp{$MAPLOOP}, match={XXX}, replace={Avtom. Vychisl. Tekh.}]
			\step[fieldsource=\regexp{$MAPLOOP}, match={XXX}, replace={Avtomatika}]
			\step[fieldsource=\regexp{$MAPLOOP}, match={XXX}, replace={Avtometriya}]
			\step[fieldsource=\regexp{$MAPLOOP}, match={ATZelektronik worldwide}, replace={ATZelektron. worldw.}]
			\step[fieldsource=\regexp{$MAPLOOP}, match={XXX}, replace={Ber. Bunsenges. Phys. Chem.}]
			\step[fieldsource=\regexp{$MAPLOOP}, match={XXX}, replace={Biofizika}]
			\step[fieldsource=\regexp{$MAPLOOP}, match={XXX}, replace={Biometrika}]
			\step[fieldsource=\regexp{$MAPLOOP}, match={XXX}, replace={Boll. Geofis. Teor. Appl.}]
			\step[fieldsource=\regexp{$MAPLOOP}, match={XXX}, replace={Bull. Acad. Serbe Sci. Arts cl. Sci. Tech.}]
			\step[fieldsource=\regexp{$MAPLOOP}, match={XXX}, replace={Bull. Annu. Soc. Suisse Chronom. Lab. Suisse.}]
			\step[fieldsource=\regexp{$MAPLOOP}, match={XXX}, replace={Rech. Horlog.}]
			\step[fieldsource=\regexp{$MAPLOOP}, match={XXX}, replace={Bull. Cl. Sci. Acad. R. Belg.}]
			\step[fieldsource=\regexp{$MAPLOOP}, match={XXX}, replace={Bull. Dir. Etud. Rech. A}]
			\step[fieldsource=\regexp{$MAPLOOP}, match={XXX}, replace={Bull. Dir. Etud. Rech. B}]
			\step[fieldsource=\regexp{$MAPLOOP}, match={XXX}, replace={Bull. Dir. Etud. Rech. C}]
			\step[fieldsource=\regexp{$MAPLOOP}, match={XXX}, replace={Bull. Liaison Rech. Inform. Autom.}]
			\step[fieldsource=\regexp{$MAPLOOP}, match={XXX}, replace={Bur. Etud. Autom.}]
			\step[fieldsource=\regexp{$MAPLOOP}, match={XXX}, replace={CFI-Ceram. Forum Int.-Ber. Dtsch. Keram. Ges.}]
			\step[fieldsource=\regexp{$MAPLOOP}, match={XXX}, replace={Chem. Scr.}]
			\step[fieldsource=\regexp{$MAPLOOP}, match={XXX}, replace={Ciel Terre}]
			\step[fieldsource=\regexp{$MAPLOOP}, match={XXX}, replace={Cybernetica}]
			\step[fieldsource=\regexp{$MAPLOOP}, match={XXX}, replace={Deut. Hydrogr. Z.}]
			\step[fieldsource=\regexp{$MAPLOOP}, match={XXX}, replace={Dokl. Akad. Nauk SSSR}]
			\step[fieldsource=\regexp{$MAPLOOP}, match={XXX}, replace={Electroacoustique}]
			\step[fieldsource=\regexp{$MAPLOOP}, match={XXX}, replace={Electrochim. Acta}]
			\step[fieldsource=\regexp{$MAPLOOP}, match={XXX}, replace={Electrochim. Metal.}]
			\step[fieldsource=\regexp{$MAPLOOP}, match={XXX}, replace={Elektor Electron.}]
			\step[fieldsource=\regexp{$MAPLOOP}, match={XXX}, replace={Elektr. Bahnen}]
			\step[fieldsource=\regexp{$MAPLOOP}, match={XXX}, replace={Elektr. Energ.-Tech.}]
			\step[fieldsource=\regexp{$MAPLOOP}, match={XXX}, replace={Elektr. Masch.}]
			\step[fieldsource=\regexp{$MAPLOOP}, match={XXX}, replace={Elektr. Stn.}]
			\step[fieldsource=\regexp{$MAPLOOP}, match={XXX}, replace={Elektrichestvo}]
			\step[fieldsource=\regexp{$MAPLOOP}, match={XXX}, replace={Elektrie}]
			\step[fieldsource=\regexp{$MAPLOOP}, match={XXX}, replace={Elektrizitaetswirtschaft}]
			\step[fieldsource=\regexp{$MAPLOOP}, match={XXX}, replace={Elektro}]
			\step[fieldsource=\regexp{$MAPLOOP}, match={XXX}, replace={Elektro-Anz.}]
			\step[fieldsource=\regexp{$MAPLOOP}, match={XXX}, replace={Elektro-Jahr}]
			\step[fieldsource=\regexp{$MAPLOOP}, match={XXX}, replace={Elektrokhimiya}]
			\step[fieldsource=\regexp{$MAPLOOP}, match={XXX}, replace={Elektron}]
			\step[fieldsource=\regexp{$MAPLOOP}, match={XXX}, replace={Elektron Int.}]
			\step[fieldsource=\regexp{$MAPLOOP}, match={XXX}, replace={Elektron. Entwitkl.}]
			\step[fieldsource=\regexp{$MAPLOOP}, match={XXX}, replace={Elektron. Ind}]
			\step[fieldsource=\regexp{$MAPLOOP}, match={XXX}, replace={Elektron. J.}]
			\step[fieldsource=\regexp{$MAPLOOP}, match={XXX}, replace={Elektron. Prax.}]
			\step[fieldsource=\regexp{$MAPLOOP}, match={XXX}, replace={Elektron. Tekh.}]
			\step[fieldsource=\regexp{$MAPLOOP}, match={XXX}, replace={Elektronica}]
			\step[fieldsource=\regexp{$MAPLOOP}, match={XXX}, replace={Elektronik}]
			\step[fieldsource=\regexp{$MAPLOOP}, match={XXX}, replace={Elektronika}]
			\step[fieldsource=\regexp{$MAPLOOP}, match={XXX}, replace={Elektroniker}]
			\step[fieldsource=\regexp{$MAPLOOP}, match={XXX}, replace={Elektronikschau}]
			\step[fieldsource=\regexp{$MAPLOOP}, match={XXX}, replace={Elektrosvyaz}]
			\step[fieldsource=\regexp{$MAPLOOP}, match={XXX}, replace={Elektrotech. Cas.}]
			\step[fieldsource=\regexp{$MAPLOOP}, match={Elektrotechnik und Informationstechnik}, replace={Elektrotech. Inf. Tech.}]
			\step[fieldsource=\regexp{$MAPLOOP}, match={XXX}, replace={Elektrotech. Obz.}]
			\step[fieldsource=\regexp{$MAPLOOP}, match={XXX}, replace={Elektrotechniek}]
			\step[fieldsource=\regexp{$MAPLOOP}, match={XXX}, replace={Elektrotechnik (Czechoslovakia)}]
			\step[fieldsource=\regexp{$MAPLOOP}, match={XXX}, replace={Elektrotechnik (Switzerland)}]
			\step[fieldsource=\regexp{$MAPLOOP}, match={XXX}, replace={Elektrotechnik (Germany)}]
			\step[fieldsource=\regexp{$MAPLOOP}, match={XXX}, replace={Elektrotechnika}]
			\step[fieldsource=\regexp{$MAPLOOP}, match={XXX}, replace={Elektrotehnika, Zagreb}]
			\step[fieldsource=\regexp{$MAPLOOP}, match={XXX}, replace={Elektrotekhnika}]
			\step[fieldsource=\regexp{$MAPLOOP}, match={XXX}, replace={Elektroteknikeren}]
			\step[fieldsource=\regexp{$MAPLOOP}, match={XXX}, replace={Elektrowaerme Int. B.}]
			\step[fieldsource=\regexp{$MAPLOOP}, match={XXX}, replace={Elettrificazione}]
			\step[fieldsource=\regexp{$MAPLOOP}, match={XXX}, replace={Elettron. Oggi}]
			\step[fieldsource=\regexp{$MAPLOOP}, match={XXX}, replace={Elettron. Telecomun.}]
			\step[fieldsource=\regexp{$MAPLOOP}, match={XXX}, replace={Elettrotecnica}]
			\step[fieldsource=\regexp{$MAPLOOP}, match={XXX}, replace={Elek. Med Aktuell Elektron.}]
			\step[fieldsource=\regexp{$MAPLOOP}, match={XXX}, replace={Elteknik}]
			\step[fieldsource=\regexp{$MAPLOOP}, match={XXX}, replace={Energ. Atomtech.}]
			\step[fieldsource=\regexp{$MAPLOOP}, match={XXX}, replace={Energ. Elettr.}]
			\step[fieldsource=\regexp{$MAPLOOP}, match={XXX}, replace={Energetica}]
			\step[fieldsource=\regexp{$MAPLOOP}, match={XXX}, replace={Energetik}]
			\step[fieldsource=\regexp{$MAPLOOP}, match={XXX}, replace={Energetika}]
			\step[fieldsource=\regexp{$MAPLOOP}, match={XXX}, replace={Energetyka}]
			\step[fieldsource=\regexp{$MAPLOOP}, match={XXX}, replace={Energia Nuclear}]
			\step[fieldsource=\regexp{$MAPLOOP}, match={XXX}, replace={Energie Technik (Germany)}]
			\step[fieldsource=\regexp{$MAPLOOP}, match={XXX}, replace={Energie Technik (Switzerland)}]
			\step[fieldsource=\regexp{$MAPLOOP}, match={XXX}, replace={Entropie}]
			\step[fieldsource=\regexp{$MAPLOOP}, match={XXX}, replace={ETZ}]
			\step[fieldsource=\regexp{$MAPLOOP}, match={XXX}, replace={ETZ Arch.}]
			\step[fieldsource=\regexp{$MAPLOOP}, match={XXX}, replace={Feingeraetetechnik}]
			\step[fieldsource=\regexp{$MAPLOOP}, match={XXX}, replace={Feinw. Tech. Messtech.}]
			\step[fieldsource=\regexp{$MAPLOOP}, match={XXX}, replace={Fert. Tech. Betr.}]
			\step[fieldsource=\regexp{$MAPLOOP}, match={XXX}, replace={Fis. Tecnol.}]
			\step[fieldsource=\regexp{$MAPLOOP}, match={XXX}, replace={Fiz. Khim. Obrab. Mater.}]
			\step[fieldsource=\regexp{$MAPLOOP}, match={XXX}, replace={Fiz. Met. Metalloved}]
			\step[fieldsource=\regexp{$MAPLOOP}, match={XXX}, replace={Fiz. Nizk. Temp.}]
			\step[fieldsource=\regexp{$MAPLOOP}, match={XXX}, replace={Fiz. Plazmy}]
			\step[fieldsource=\regexp{$MAPLOOP}, match={XXX}, replace={Fiz. Tekh. Poluprovodn.}]
			\step[fieldsource=\regexp{$MAPLOOP}, match={XXX}, replace={Fiz. Tverd. Tela}]
			\step[fieldsource=\regexp{$MAPLOOP}, match={XXX}, replace={Fiz.-Khim. Mekh. Mater.}]
			\step[fieldsource=\regexp{$MAPLOOP}, match={XXX}, replace={Fizika}]
			\step[fieldsource=\regexp{$MAPLOOP}, match={XXX}, replace={Forsch.-Ber. Landes Nordrh.-Westfal.}]
			\step[fieldsource=\regexp{$MAPLOOP}, match={XXX}, replace={Frequenz}]
			\step[fieldsource=\regexp{$MAPLOOP}, match={XXX}, replace={Fys. Tidsskr.}]
			\step[fieldsource=\regexp{$MAPLOOP}, match={XXX}, replace={G. Fis.}]
			\step[fieldsource=\regexp{$MAPLOOP}, match={XXX}, replace={Geliotekhnika}]
			\step[fieldsource=\regexp{$MAPLOOP}, match={XXX}, replace={Geochim. Cosmochim. Acta}]
			\step[fieldsource=\regexp{$MAPLOOP}, match={XXX}, replace={Haerterei-Tech. Mitt.}]
			\step[fieldsource=\regexp{$MAPLOOP}, match={XXX}, replace={Helv. Chim. Acta}]
			\step[fieldsource=\regexp{$MAPLOOP}, match={XXX}, replace={Helv. Med. Acta}]
			\step[fieldsource=\regexp{$MAPLOOP}, match={XXX}, replace={Helv. Phys. Acta}]
			\step[fieldsource=\regexp{$MAPLOOP}, match={XXX}, replace={Hochfreq. Electroakust}]
			\step[fieldsource=\regexp{$MAPLOOP}, match={XXX}, replace={Hoppe-Seylers Z. Physiol. Chem.}]
			\step[fieldsource=\regexp{$MAPLOOP}, match={XXX}, replace={Inf. Elektron.}]
			\step[fieldsource=\regexp{$MAPLOOP}, match={XXX}, replace={Inf. Elettron.}]
			\step[fieldsource=\regexp{$MAPLOOP}, match={XXX}, replace={Inform. Forsch. Entwickl.}]
			\step[fieldsource=\regexp{$MAPLOOP}, match={XXX}, replace={Inform. Spektrum}]
			\step[fieldsource=\regexp{$MAPLOOP}, match={XXX}, replace={Inform.-Fachber.}]
			\step[fieldsource=\regexp{$MAPLOOP}, match={XXX}, replace={Informatie}]
			\step[fieldsource=\regexp{$MAPLOOP}, match={XXX}, replace={Informatik}]
			\step[fieldsource=\regexp{$MAPLOOP}, match={XXX}, replace={Informatologia Yugosl.}]
			\step[fieldsource=\regexp{$MAPLOOP}, match={XXX}, replace={Informatyka}]
			\step[fieldsource=\regexp{$MAPLOOP}, match={XXX}, replace={Infowelt}]
			\step[fieldsource=\regexp{$MAPLOOP}, match={XXX}, replace={Ing. Electr. Mec.}]
			\step[fieldsource=\regexp{$MAPLOOP}, match={XXX}, replace={Ing. Mec. Electr.}]
			\step[fieldsource=\regexp{$MAPLOOP}, match={XXX}, replace={Ing.-Arch.}]
			\step[fieldsource=\regexp{$MAPLOOP}, match={XXX}, replace={Inzh.-Fiz. Zh.}]
			\step[fieldsource=\regexp{$MAPLOOP}, match={XXX}, replace={Izmer. Tekh.}]
			\step[fieldsource=\regexp{$MAPLOOP}, match={XXX}, replace={Izv. Akad. Nauk Arm. SSR Ser. Tekh. Nauk}]
			\step[fieldsource=\regexp{$MAPLOOP}, match={XXX}, replace={Izv. Akad. Nauk SSSR Energ. Transp.}]
			\step[fieldsource=\regexp{$MAPLOOP}, match={XXX}, replace={Izv. Akad. Nauk SSSR Fiz. Atmos. Okeana}]
			\step[fieldsource=\regexp{$MAPLOOP}, match={XXX}, replace={Izv. Akad. Nauk SSSR Fiz. Zemli}]
			\step[fieldsource=\regexp{$MAPLOOP}, match={XXX}, replace={Izv. Akad. Nauk SSSR Ser. Fiz.}]
			\step[fieldsource=\regexp{$MAPLOOP}, match={XXX}, replace={Izv. Vyssh. Uchebn. Zaved. Elektromekh.}]
			\step[fieldsource=\regexp{$MAPLOOP}, match={XXX}, replace={Izv. Vyssh. Uchebn. Zaved. Radioelektron.}]
			\step[fieldsource=\regexp{$MAPLOOP}, match={XXX}, replace={Izv. Vyssh. Uchebn. Zaved. Radiofiz.}]
			\step[fieldsource=\regexp{$MAPLOOP}, match={XXX}, replace={J. Chim Phys. Phys.-Chim Biol.}]
			\step[fieldsource=\regexp{$MAPLOOP}, match={XXX}, replace={Kernenergie}]
			\step[fieldsource=\regexp{$MAPLOOP}, match={XXX}, replace={Kerntechnik}]
			\step[fieldsource=\regexp{$MAPLOOP}, match={XXX}, replace={Khim. Fiz.}]
			\step[fieldsource=\regexp{$MAPLOOP}, match={XXX}, replace={Kibern. Vychisl. Tekh.}]
			\step[fieldsource=\regexp{$MAPLOOP}, match={XXX}, replace={Kibernetika}]
			\step[fieldsource=\regexp{$MAPLOOP}, match={XXX}, replace={Kristallografiya}]
			\step[fieldsource=\regexp{$MAPLOOP}, match={XXX}, replace={Kvantovaya Elektron. Mosk.}]
			\step[fieldsource=\regexp{$MAPLOOP}, match={XXX}, replace={Kybernetes}]
			\step[fieldsource=\regexp{$MAPLOOP}, match={XXX}, replace={Kybernetika}]
			\step[fieldsource=\regexp{$MAPLOOP}, match={XXX}, replace={Med. Tek.}]
			\step[fieldsource=\regexp{$MAPLOOP}, match={XXX}, replace={Mekh. Avtom. Proizvod.}]
			\step[fieldsource=\regexp{$MAPLOOP}, match={XXX}, replace={Meres Autom.}]
			\step[fieldsource=\regexp{$MAPLOOP}, match={XXX}, replace={Mesures}]
			\step[fieldsource=\regexp{$MAPLOOP}, match={XXX}, replace={Metallofizika}]
			\step[fieldsource=\regexp{$MAPLOOP}, match={XXX}, replace={Metalloved. Term. Obrab. Met.}]
			\step[fieldsource=\regexp{$MAPLOOP}, match={XXX}, replace={Meteorol. Gidrol.}]
			\step[fieldsource=\regexp{$MAPLOOP}, match={XXX}, replace={Meteorol. Rundsch.}]
			\step[fieldsource=\regexp{$MAPLOOP}, match={XXX}, replace={Metrol. Apl.}]
			\step[fieldsource=\regexp{$MAPLOOP}, match={XXX}, replace={Metrologia}]
			\step[fieldsource=\regexp{$MAPLOOP}, match={XXX}, replace={Medel. Simul.}]
			\step[fieldsource=\regexp{$MAPLOOP}, match={XXX}, replace={Nachr. Dok.}]
			\step[fieldsource=\regexp{$MAPLOOP}, match={XXX}, replace={Nachr.tech. Elektron.}]
			\step[fieldsource=\regexp{$MAPLOOP}, match={XXX}, replace={Naturwissentchaften}]
			\step[fieldsource=\regexp{$MAPLOOP}, match={XXX}, replace={Neue Tech.}]
			\step[fieldsource=\regexp{$MAPLOOP}, match={XXX}, replace={Neue Tech. Buero}]
			\step[fieldsource=\regexp{$MAPLOOP}, match={XXX}, replace={Nukleonika}]
			\step[fieldsource=\regexp{$MAPLOOP}, match={XXX}, replace={Numer. Math.}]
			\step[fieldsource=\regexp{$MAPLOOP}, match={XXX}, replace={Nuovo Cimento A}]
			\step[fieldsource=\regexp{$MAPLOOP}, match={XXX}, replace={Nuovo Cimento B}]
			\step[fieldsource=\regexp{$MAPLOOP}, match={XXX}, replace={Nouvo Cimento C}]
			\step[fieldsource=\regexp{$MAPLOOP}, match={XXX}, replace={Nuovo Cimento D}]
			\step[fieldsource=\regexp{$MAPLOOP}, match={XXX}, replace={Okeanologiya}]
			\step[fieldsource=\regexp{$MAPLOOP}, match={XXX}, replace={Opt. Spektrosk.}]
			\step[fieldsource=\regexp{$MAPLOOP}, match={XXX}, replace={Opt.-Mekh. Prom.}]
			\step[fieldsource=\regexp{$MAPLOOP}, match={XXX}, replace={Optik}]
			\step[fieldsource=\regexp{$MAPLOOP}, match={XXX}, replace={Photogrammetria}]
			\step[fieldsource=\regexp{$MAPLOOP}, match={XXX}, replace={Photonics Spectra}]
			\step[fieldsource=\regexp{$MAPLOOP}, match={XXX}, replace={Pis’ma Astron. Zh. }]
			\step[fieldsource=\regexp{$MAPLOOP}, match={XXX}, replace={Pis’ma Zh. Eksp. Teor. Fiz. }]
			\step[fieldsource=\regexp{$MAPLOOP}, match={XXX}, replace={Pis’ma Zh. Tekh. Fiz. }]
			\step[fieldsource=\regexp{$MAPLOOP}, match={XXX}, replace={Poverkhn., Fiz. Khim. Mekh.}]
			\step[fieldsource=\regexp{$MAPLOOP}, match={XXX}, replace={Pr. Inst. Elektrotech.}]
			\step[fieldsource=\regexp{$MAPLOOP}, match={XXX}, replace={Prib. Sist. Upr.}]
			\step[fieldsource=\regexp{$MAPLOOP}, match={XXX}, replace={Prib. Tekh. Eksp.}]
			\step[fieldsource=\regexp{$MAPLOOP}, match={XXX}, replace={Prikl. Mat. Mekh.}]
			\step[fieldsource=\regexp{$MAPLOOP}, match={XXX}, replace={Prikl. Mekh.}]
			\step[fieldsource=\regexp{$MAPLOOP}, match={XXX}, replace={Probl. Kibern.}]
			\step[fieldsource=\regexp{$MAPLOOP}, match={XXX}, replace={Probl. Peredachi Inf.}]
			\step[fieldsource=\regexp{$MAPLOOP}, match={XXX}, replace={Proc. K. Ned. Akad. Wet. B, Palaeontol.}]
			\step[fieldsource=\regexp{$MAPLOOP}, match={XXX}, replace={Anthropol. }]
			\step[fieldsource=\regexp{$MAPLOOP}, match={XXX}, replace={Programmirovanie}]
			\step[fieldsource=\regexp{$MAPLOOP}, match={XXX}, replace={Prz. Elektrotech.}]
			\step[fieldsource=\regexp{$MAPLOOP}, match={XXX}, replace={Prz. Telekomun.}]
			\step[fieldsource=\regexp{$MAPLOOP}, match={XXX}, replace={PT/Elektrotech. Elektron.}]
			\step[fieldsource=\regexp{$MAPLOOP}, match={XXX}, replace={Radio Fernsehen Elektron.}]
			\step[fieldsource=\regexp{$MAPLOOP}, match={XXX}, replace={Radiotekh. Elektron.}]
			\step[fieldsource=\regexp{$MAPLOOP}, match={XXX}, replace={Radiotekhnika Mosk.}]
			\step[fieldsource=\regexp{$MAPLOOP}, match={XXX}, replace={Rev. Acad. Cienc. Zaragoza}]
			\step[fieldsource=\regexp{$MAPLOOP}, match={XXX}, replace={Rev. Electrotec. (Argentina)}]
			\step[fieldsource=\regexp{$MAPLOOP}, match={XXX}, replace={Rev. Electrotec. (Spain)}]
			\step[fieldsource=\regexp{$MAPLOOP}, match={XXX}, replace={Rev. Energ.}]
			\step[fieldsource=\regexp{$MAPLOOP}, match={XXX}, replace={Rev. Esp. Electron. Rev. Geofis.}]
			\step[fieldsource=\regexp{$MAPLOOP}, match={XXX}, replace={Ric. Autom.}]
			\step[fieldsource=\regexp{$MAPLOOP}, match={XXX}, replace={Ric. Spettrosc.}]
			\step[fieldsource=\regexp{$MAPLOOP}, match={XXX}, replace={Robotersysteme}]
			\step[fieldsource=\regexp{$MAPLOOP}, match={XXX}, replace={Rozpr. Electrotech.}]
			\step[fieldsource=\regexp{$MAPLOOP}, match={XXX}, replace={Sadhana}]
			\step[fieldsource=\regexp{$MAPLOOP}, match={XXX}, replace={Schweiz. Tech. Z.}]
			\step[fieldsource=\regexp{$MAPLOOP}, match={XXX}, replace={Scientia}]
			\step[fieldsource=\regexp{$MAPLOOP}, match={XXX}, replace={Siemens Forsch. Entwickl. Ber.}]
			\step[fieldsource=\regexp{$MAPLOOP}, match={XXX}, replace={Sist. Autom.}]
			\step[fieldsource=\regexp{$MAPLOOP}, match={XXX}, replace={Sitzungsber. Oester. Akad. Wiss. Math.-}]
			\step[fieldsource=\regexp{$MAPLOOP}, match={XXX}, replace={Naturwiss. Kl. Abt. II (Austria)}]
			\step[fieldsource=\regexp{$MAPLOOP}, match={XXX}, replace={Spectrochim. Acta A, Mol. Spectrosc.}]
			\step[fieldsource=\regexp{$MAPLOOP}, match={XXX}, replace={Spectrochim. Acta B, At. Spectrosc.}]
			\step[fieldsource=\regexp{$MAPLOOP}, match={XXX}, replace={Sprache Datenverarb.}]
			\step[fieldsource=\regexp{$MAPLOOP}, match={XXX}, replace={Stanki Instrum.}]
			\step[fieldsource=\regexp{$MAPLOOP}, match={XXX}, replace={Steklo Keram.}]
			\step[fieldsource=\regexp{$MAPLOOP}, match={XXX}, replace={Svetotekhnika  TE Int.}]
			\step[fieldsource=\regexp{$MAPLOOP}, match={XXX}, replace={Tech. Bull. Vevey}]
			\step[fieldsource=\regexp{$MAPLOOP}, match={XXX}, replace={Tech. Mitt. Krupp (Engl. Ed.)}]
			\step[fieldsource=\regexp{$MAPLOOP}, match={XXX}, replace={Tech. Mitt. PTT}]
			\step[fieldsource=\regexp{$MAPLOOP}, match={XXX}, replace={Tech. Mitt. RFZ}]
			\step[fieldsource=\regexp{$MAPLOOP}, match={XXX}, replace={Technica}]
			\step[fieldsource=\regexp{$MAPLOOP}, match={XXX}, replace={Tecnica}]
			\step[fieldsource=\regexp{$MAPLOOP}, match={XXX}, replace={Teh. Fiz.}]
			\step[fieldsource=\regexp{$MAPLOOP}, match={XXX}, replace={Tehnika}]
			\step[fieldsource=\regexp{$MAPLOOP}, match={XXX}, replace={Tekh. Elektrodin.}]
			\step[fieldsource=\regexp{$MAPLOOP}, match={XXX}, replace={Tekh. Kibern.}]
			\step[fieldsource=\regexp{$MAPLOOP}, match={XXX}, replace={Tekh. Kino Telev.}]
			\step[fieldsource=\regexp{$MAPLOOP}, match={XXX}, replace={Tekh. Misul}]
			\step[fieldsource=\regexp{$MAPLOOP}, match={XXX}, replace={Telekomunikacije}]
			\step[fieldsource=\regexp{$MAPLOOP}, match={XXX}, replace={Telektronikk}]
			\step[fieldsource=\regexp{$MAPLOOP}, match={XXX}, replace={Teleteknik}]
			\step[fieldsource=\regexp{$MAPLOOP}, match={XXX}, replace={Teor. Mat. Fiz.}]
			\step[fieldsource=\regexp{$MAPLOOP}, match={XXX}, replace={Teploeoergetika}]
			\step[fieldsource=\regexp{$MAPLOOP}, match={XXX}, replace={Teplofiz. Vvs. Temp.}]
			\step[fieldsource=\regexp{$MAPLOOP}, match={XXX}, replace={Tidskr. Dok.}]
			\step[fieldsource=\regexp{$MAPLOOP}, match={XXX}, replace={TN Nachr.}]
			\step[fieldsource=\regexp{$MAPLOOP}, match={XXX}, replace={Toute Electron.}]
			\step[fieldsource=\regexp{$MAPLOOP}, match={XXX}, replace={Tr. Inst. Teor. Astron.}]
			\step[fieldsource=\regexp{$MAPLOOP}, match={XXX}, replace={Ukr. Fiz. Zh.}]
			\step[fieldsource=\regexp{$MAPLOOP}, match={XXX}, replace={Usp. Fiz. Nauk}]
			\step[fieldsource=\regexp{$MAPLOOP}, match={XXX}, replace={Vak.-Tech.}]
			\step[fieldsource=\regexp{$MAPLOOP}, match={XXX}, replace={VDE Fachiber.}]
			\step[fieldsource=\regexp{$MAPLOOP}, match={XXX}, replace={VDI Z.}]
			\step[fieldsource=\regexp{$MAPLOOP}, match={XXX}, replace={Vestn. Mashinostr.}]
			\step[fieldsource=\regexp{$MAPLOOP}, match={XXX}, replace={Vestn. Mosk. Univ. 15, Vychisl. Mat. Kibern.}]
			\step[fieldsource=\regexp{$MAPLOOP}, match={XXX}, replace={Vestn. Mosk. Univ. 3, Fiz. Astron.}]
			\step[fieldsource=\regexp{$MAPLOOP}, match={XXX}, replace={Vesti Akad. Navuk BSSR Ser. Fiz. Energ. Navuk}]
			\step[fieldsource=\regexp{$MAPLOOP}, match={XXX}, replace={VGB Kraftwerkstech. (Ger. Ed.)}]
			\step[fieldsource=\regexp{$MAPLOOP}, match={XXX}, replace={Vide Couches Minces}]
			\step[fieldsource=\regexp{$MAPLOOP}, match={XXX}, replace={Vistas Astron.}]
			\step[fieldsource=\regexp{$MAPLOOP}, match={XXX}, replace={Vopr. At. Nauki Tekh. Ser., Fiz. Radiats.}]
			\step[fieldsource=\regexp{$MAPLOOP}, match={XXX}, replace={Povrezhdenii Radiats. Materialoved.}]
			\step[fieldsource=\regexp{$MAPLOOP}, match={XXX}, replace={Vopr. At. Nauki Tekh. Ser., Obshch. Yad. Fiz.}]
			\step[fieldsource=\regexp{$MAPLOOP}, match={XXX}, replace={Vuoto Sci. Tecnol.}]
			\step[fieldsource=\regexp{$MAPLOOP}, match={XXX}, replace={Wiss. Z. Friedrich-Schiller-Univ. Jena}]
			\step[fieldsource=\regexp{$MAPLOOP}, match={XXX}, replace={Nat.wiss. Reihe}]
			\step[fieldsource=\regexp{$MAPLOOP}, match={XXX}, replace={Wiss. Z. Karl-Marx-Univ. Leipz. Math.-}]
			\step[fieldsource=\regexp{$MAPLOOP}, match={XXX}, replace={Nat.wiss. Reihe}]
			\step[fieldsource=\regexp{$MAPLOOP}, match={XXX}, replace={Wiss. Z. Tech. Hochsch. Ilmenau}]
			\step[fieldsource=\regexp{$MAPLOOP}, match={XXX}, replace={Wiss. Z. Tech. Univ. Dresd.}]
			\step[fieldsource=\regexp{$MAPLOOP}, match={XXX}, replace={Wiss. Z. Tech. Univ. Karl-Marx-Stadt}]
			\step[fieldsource=\regexp{$MAPLOOP}, match={XXX}, replace={Yad. Fiz.}]
			\step[fieldsource=\regexp{$MAPLOOP}, match={XXX}, replace={Z. Angew. Math. Mech.}]
			\step[fieldsource=\regexp{$MAPLOOP}, match={XXX}, replace={Z. Angew. Math. Phys.}]
			\step[fieldsource=\regexp{$MAPLOOP}, match={XXX}, replace={Z. Met.kd.}]
			\step[fieldsource=\regexp{$MAPLOOP}, match={XXX}, replace={Z. Nat. Forsch. A, Phys. Phys. Chem. Kosmophys.}]
			\step[fieldsource=\regexp{$MAPLOOP}, match={XXX}, replace={Z. Oper. Res. A, Theor.}]
			\step[fieldsource=\regexp{$MAPLOOP}, match={XXX}, replace={Z. Oper. Res. B, Prax.}]
			\step[fieldsource=\regexp{$MAPLOOP}, match={XXX}, replace={Z. Phys.}]
			\step[fieldsource=\regexp{$MAPLOOP}, match={XXX}, replace={Z. Phys. A, At. Nuclei}]
			\step[fieldsource=\regexp{$MAPLOOP}, match={XXX}, replace={Z. Phys. B, Condens. Matter}]
			\step[fieldsource=\regexp{$MAPLOOP}, match={XXX}, replace={Z. Phys. C, Part Fields}]
			\step[fieldsource=\regexp{$MAPLOOP}, match={XXX}, replace={Z. Phys. Chem. Neue Folge}]
			\step[fieldsource=\regexp{$MAPLOOP}, match={XXX}, replace={Z. Phys. Chem., Leipz.}]
			\step[fieldsource=\regexp{$MAPLOOP}, match={XXX}, replace={Z. Phys. D, At. Mol. Clusters}]
			\step[fieldsource=\regexp{$MAPLOOP}, match={XXX}, replace={Zavod. Lab.}]
			\step[fieldsource=\regexp{$MAPLOOP}, match={XXX}, replace={Zh. Eksp. Teor. Fiz.}]
			\step[fieldsource=\regexp{$MAPLOOP}, match={XXX}, replace={Zh. Fiz. Khim.}]
			\step[fieldsource=\regexp{$MAPLOOP}, match={XXX}, replace={Zh. Prikl. Mekh. Tekh. Fiz.}]
			\step[fieldsource=\regexp{$MAPLOOP}, match={XXX}, replace={Zh. Prikl. Spektrosk.}]
			\step[fieldsource=\regexp{$MAPLOOP}, match={XXX}, replace={Zh. Tekh. Fiz.}]
			\step[fieldsource=\regexp{$MAPLOOP}, match={XXX}, replace={Zh. Vychisl. Mat. Mat. Fiz.}]
			\step[fieldsource=\regexp{$MAPLOOP}, match={XXX}, replace={Zisin, J. Seismol. Soc. Jpn.}] 
			\step[fieldsource=\regexp{$MAPLOOP}, match={Production}, replace={Prod.}]
			\step[fieldsource=\regexp{$MAPLOOP}, match={Reliability}, replace={Rel.}]
			\step[fieldsource=\regexp{$MAPLOOP}, match={Report}, replace={Rep.}]
			\step[fieldsource=\regexp{$MAPLOOP}, match={Semiconductor}, replace={Semicond.}]
			\step[fieldsource=\regexp{$MAPLOOP}, match={Research}, replace={Res.}]
			\step[fieldsource=\regexp{$MAPLOOP}, match={Sensing}, replace={Sens.}]
			\step[fieldsource=\regexp{$MAPLOOP}, match={Resonance}, replace={Reson.}]
			\step[fieldsource=\regexp{$MAPLOOP}, match={Series}, replace={Ser.}]
			\step[fieldsource=\regexp{$MAPLOOP}, match={Resources}, replace={Resour.}]
			\step[fieldsource=\regexp{$MAPLOOP}, match={Simulation}, replace={Simul.}]
			\step[fieldsource=\regexp{$MAPLOOP}, match={Reviews}, replace={Rev.}]
			\step[fieldsource=\regexp{$MAPLOOP}, match={Review}, replace={Rev.}]
			\step[fieldsource=\regexp{$MAPLOOP}, match={Singapore}, replace={Singap.}]
			\step[fieldsource=\regexp{$MAPLOOP}, match={Robotics}, replace={Robot.}]
			\step[fieldsource=\regexp{$MAPLOOP}, match={Sistema}, replace={Sist.}]
			\step[fieldsource=\regexp{$MAPLOOP}, match={Royal}, replace={Roy.}]
			\step[fieldsource=\regexp{$MAPLOOP}, match={Society}, replace={Soc.}]
			\step[fieldsource=\regexp{$MAPLOOP}, match={Safety}, replace={Saf.}]
			\step[fieldsource=\regexp{$MAPLOOP}, match={Sociological}, replace={Sociol.}]
			\step[fieldsource=\regexp{$MAPLOOP}, match={Satellite}, replace={Satell.}]
			\step[fieldsource=\regexp{$MAPLOOP}, match={Software}, replace={Softw.}]
			\step[fieldsource=\regexp{$MAPLOOP}, match={Scandinavian}, replace={Scand.}]
			\step[fieldsource=\regexp{$MAPLOOP}, match={Solar}, replace={Sol.}]
			\step[fieldsource=\regexp{$MAPLOOP}, match={Sciences}, replace={Sci.}]
			\step[fieldsource=\regexp{$MAPLOOP}, match={Science}, replace={Sci.}]
			\step[fieldsource=\regexp{$MAPLOOP}, match={Soviet}, replace={Sov.}]
			\step[fieldsource=\regexp{$MAPLOOP}, match={Section}, replace={Sect.}]
			\step[fieldsource=\regexp{$MAPLOOP}, match={Spectroscopy}, replace={Spectrosc.}]
			\step[fieldsource=\regexp{$MAPLOOP}, match={Security}, replace={Secur.}]
			\step[fieldsource=\regexp{$MAPLOOP}, match={Spectrum}, replace={Spectr.}]
			\step[fieldsource=\regexp{$MAPLOOP}, match={Seismology}, replace={Seismol.}]
			\step[fieldsource=\regexp{$MAPLOOP}, match={Speculations}, replace={Specul.}]
			\step[fieldsource=\regexp{$MAPLOOP}, match={Selected}, replace={Sel.}]
			\step[fieldsource=\regexp{$MAPLOOP}, match={Statistics}, replace={Statist.}]
			\step[fieldsource=\regexp{$MAPLOOP}, match={Structures}, replace={Struct.}]
			\step[fieldsource=\regexp{$MAPLOOP}, match={Structure}, replace={Struct.}]
			\step[fieldsource=\regexp{$MAPLOOP}, match={Terrestrial}, replace={Terr.}]
			\step[fieldsource=\regexp{$MAPLOOP}, match={Studies}, replace={Stud.}]
			\step[fieldsource=\regexp{$MAPLOOP}, match={Theoretical}, replace={Theor.}]
			\step[fieldsource=\regexp{$MAPLOOP}, match={Superconductivity}, replace={Supercond.}]
			\step[fieldsource=\regexp{$MAPLOOP}, match={Transactions}, replace={Trans.}]
			\step[fieldsource=\regexp{$MAPLOOP}, match={Supplement}, replace={Suppl.}]
			\step[fieldsource=\regexp{$MAPLOOP}, match={Translation}, replace={Transl.}]
			\step[fieldsource=\regexp{$MAPLOOP}, match={Surface}, replace={Surf.}]
			\step[fieldsource=\regexp{$MAPLOOP}, match={Transmission}, replace={Transmiss.}]
			\step[fieldsource=\regexp{$MAPLOOP}, match={Survey}, replace={Surv.}]
			\step[fieldsource=\regexp{$MAPLOOP}, match={Transportation}, replace={Transp.}]
			\step[fieldsource=\regexp{$MAPLOOP}, match={Sustainable}, replace={Sustain.}]
			\step[fieldsource=\regexp{$MAPLOOP}, match={Tutorials}, replace={Tut.}]
			\step[fieldsource=\regexp{$MAPLOOP}, match={Symposium}, replace={Symp.}]
			\step[fieldsource=\regexp{$MAPLOOP}, match={Ultrasonic}, replace={Ultrason.}]
			\step[fieldsource=\regexp{$MAPLOOP}, match={Systems}, replace={Syst.}]
			\step[fieldsource=\regexp{$MAPLOOP}, match={System}, replace={Syst.}]
			\step[fieldsource=\regexp{$MAPLOOP}, match={University}, replace={Univ.}]
			\step[fieldsource=\regexp{$MAPLOOP}, match={\detokenize{Universität}}, replace={Univ.}]
			\step[fieldsource=\regexp{$MAPLOOP}, match={\detokenize{Université}}, replace={Univ.}]
			\step[fieldsource=\regexp{$MAPLOOP}, match={Vacuum}, replace={Vac.}]
			\step[fieldsource=\regexp{$MAPLOOP}, match={Vehicles}, replace={Veh.}]
			\step[fieldsource=\regexp{$MAPLOOP}, match={Vehicle}, replace={Veh.}]
			\step[fieldsource=\regexp{$MAPLOOP}, match={Vehicular}, replace={Veh.}]
			\step[fieldsource=\regexp{$MAPLOOP}, match={Technology}, replace={Technol.}]
			\step[fieldsource=\regexp{$MAPLOOP}, match={Technological}, replace={Technol.}]
			\step[fieldsource=\regexp{$MAPLOOP}, match={Vibration}, replace={Vib.}]
			\step[fieldsource=\regexp{$MAPLOOP}, match={Telecommunications}, replace={Telecommun.}]
			\step[fieldsource=\regexp{$MAPLOOP}, match={Visual}, replace={Vis.}]
			\step[fieldsource=\regexp{$MAPLOOP}, match={Television}, replace={Telev.}]
			\step[fieldsource=\regexp{$MAPLOOP}, match={Welding}, replace={Weld.}]
			\step[fieldsource=\regexp{$MAPLOOP}, match={Temperature}, replace={Temp.}]
			\step[fieldsource=\regexp{$MAPLOOP}, match={Working}, replace={Work.}]
			\step[fieldsource=\regexp{$MAPLOOP}, match={Learning}, replace={Learn.}]
			\step[fieldsource=\regexp{$MAPLOOP}, match={Measurement}, replace={Meas.}]
			\step[fieldsource=\regexp{$MAPLOOP}, match={Letters}, replace={Lett.}]
			\step[fieldsource=\regexp{$MAPLOOP}, match={Letter}, replace={Lett.}]
			\step[fieldsource=\regexp{$MAPLOOP}, match={Mechanical}, replace={Mech.}]
			\step[fieldsource=\regexp{$MAPLOOP}, match={Mechanics}, replace={Mech.}]
			\step[fieldsource=\regexp{$MAPLOOP}, match={Mechanic}, replace={Mech.}]
			\step[fieldsource=\regexp{$MAPLOOP}, match={Lightwave}, replace={Lightw.}]
			\step[fieldsource=\regexp{$MAPLOOP}, match={Medical}, replace={Med.}]
			\step[fieldsource=\regexp{$MAPLOOP}, match={Logic, Logical}, replace={Log.}]
			\step[fieldsource=\regexp{$MAPLOOP}, match={Metals}, replace={Met.}]
			\step[fieldsource=\regexp{$MAPLOOP}, match={Luminescence}, replace={Lumin.}]
			\step[fieldsource=\regexp{$MAPLOOP}, match={Metallurgy}, replace={Metall.}]
			\step[fieldsource=\regexp{$MAPLOOP}, match={Machines}, replace={Mach.}]
			\step[fieldsource=\regexp{$MAPLOOP}, match={Machine}, replace={Mach.}]
			\step[fieldsource=\regexp{$MAPLOOP}, match={Meteorology}, replace={Meteorol.}]
			\step[fieldsource=\regexp{$MAPLOOP}, match={Magazine}, replace={Mag.}]
			\step[fieldsource=\regexp{$MAPLOOP}, match={Metropolitan}, replace={Metrop.}]
			\step[fieldsource=\regexp{$MAPLOOP}, match={Magnetics}, replace={Magn.}]
			\step[fieldsource=\regexp{$MAPLOOP}, match={Mexican, Mexico}, replace={Mex.}]
			\step[fieldsource=\regexp{$MAPLOOP}, match={Management}, replace={Manage.}]
			\step[fieldsource=\regexp{$MAPLOOP}, match={Microelectromechanical}, replace={Microelectromech.}]
			\step[fieldsource=\regexp{$MAPLOOP}, match={Managing}, replace={Manag.}]
			\step[fieldsource=\regexp{$MAPLOOP}, match={Microgravity}, replace={Microgr.}]
			\step[fieldsource=\regexp{$MAPLOOP}, match={Manufacturing}, replace={Manuf.}]
			\step[fieldsource=\regexp{$MAPLOOP}, match={Microscopy}, replace={Microsc.}]
			\step[fieldsource=\regexp{$MAPLOOP}, match={Marine}, replace={Mar.}]
			\step[fieldsource=\regexp{$MAPLOOP}, match={Microwaves}, replace={Microw.}]
			\step[fieldsource=\regexp{$MAPLOOP}, match={Microwave}, replace={Microw.}]
			\step[fieldsource=\regexp{$MAPLOOP}, match={Materials}, replace={Mater.}]
			\step[fieldsource=\regexp{$MAPLOOP}, match={Material}, replace={Mater.}]
			\step[fieldsource=\regexp{$MAPLOOP}, match={Military}, replace={Mil.}]
			\step[fieldsource=\regexp{$MAPLOOP}, match={Modeling}, replace={Model.}]
			\step[fieldsource=\regexp{$MAPLOOP}, match={Modelling}, replace={Model.}]
			\step[fieldsource=\regexp{$MAPLOOP}, match={Oceanic}, replace={Ocean.}]
			\step[fieldsource=\regexp{$MAPLOOP}, match={Molecular}, replace={Mol.}]
			\step[fieldsource=\regexp{$MAPLOOP}, match={Oceanography}, replace={Oceanogr.}]
			\step[fieldsource=\regexp{$MAPLOOP}, match={Monitoring}, replace={Monit.}]
			\step[fieldsource=\regexp{$MAPLOOP}, match={Occupation}, replace={Occupat.}]
			\step[fieldsource=\regexp{$MAPLOOP}, match={Multiphysics}, replace={Multiphys.}]
			\step[fieldsource=\regexp{$MAPLOOP}, match={Operational}, replace={Oper.}]
			\step[fieldsource=\regexp{$MAPLOOP}, match={Nanobioscience}, replace={Nanobiosci.}]
			\step[fieldsource=\regexp{$MAPLOOP}, match={Optical}, replace={Opt.}]
			\step[fieldsource=\regexp{$MAPLOOP}, match={Nanotechnology}, replace={Nanotechnol.}]
			\step[fieldsource=\regexp{$MAPLOOP}, match={Optics}, replace={Opt.}]
			\step[fieldsource=\regexp{$MAPLOOP}, match={National}, replace={Nat.}]
			\step[fieldsource=\regexp{$MAPLOOP}, match={Optimization}, replace={Optim.}]
			\step[fieldsource=\regexp{$MAPLOOP}, match={Naval}, replace={Nav.}]
			\step[fieldsource=\regexp{$MAPLOOP}, match={Organization}, replace={Org.}]
			\step[fieldsource=\regexp{$MAPLOOP}, match={Networking}, replace={Netw.}]
			\step[fieldsource=\regexp{$MAPLOOP}, match={Networked}, replace={Netw.}]
			\step[fieldsource=\regexp{$MAPLOOP}, match={Network}, replace={Netw.}]
			\step[fieldsource=\regexp{$MAPLOOP}, match={Packaging}, replace={Packag.}]
			\step[fieldsource=\regexp{$MAPLOOP}, match={Newsletter}, replace={Newslett.}]
			\step[fieldsource=\regexp{$MAPLOOP}, match={Particle}, replace={Part.}]
			\step[fieldsource=\regexp{$MAPLOOP}, match={Nondestructive}, replace={Nondestruct.}]
			\step[fieldsource=\regexp{$MAPLOOP}, match={Patent}, replace={Pat.}]
			\step[fieldsource=\regexp{$MAPLOOP}, match={Nuclear}, replace={Nucl.}]
			\step[fieldsource=\regexp{$MAPLOOP}, match={Performance}, replace={Perform.}]
			\step[fieldsource=\regexp{$MAPLOOP}, match={Numerical}, replace={Numer.}]
			\step[fieldsource=\regexp{$MAPLOOP}, match={Personal}, replace={Pers.}]
			\step[fieldsource=\regexp{$MAPLOOP}, match={Observations}, replace={Observ.}]
			\step[fieldsource=\regexp{$MAPLOOP}, match={Philosophical}, replace={Philos.}]
			\step[fieldsource=\regexp{$MAPLOOP}, match={Photonics}, replace={Photon.}]
			\step[fieldsource=\regexp{$MAPLOOP}, match={Productivity}, replace={Productiv.}]
			\step[fieldsource=\regexp{$MAPLOOP}, match={Photovoltaics}, replace={Photovolt.}]
			\step[fieldsource=\regexp{$MAPLOOP}, match={Programming}, replace={Program.}]
			\step[fieldsource=\regexp{$MAPLOOP}, match={Physics}, replace={Phys.}]
			\step[fieldsource=\regexp{$MAPLOOP}, match={Progress}, replace={Prog.}]
			\step[fieldsource=\regexp{$MAPLOOP}, match={Physiology}, replace={Physiol.}]
			\step[fieldsource=\regexp{$MAPLOOP}, match={Propagation}, replace={Propag.}]
			\step[fieldsource=\regexp{$MAPLOOP}, match={Planetary}, replace={Planet.}]
			\step[fieldsource=\regexp{$MAPLOOP}, match={Psychology}, replace={Psychol.}]
			\step[fieldsource=\regexp{$MAPLOOP}, match={Pneumatics}, replace={Pneum.}]
			\step[fieldsource=\regexp{$MAPLOOP}, match={Quality}, replace={Qual.}]
			\step[fieldsource=\regexp{$MAPLOOP}, match={Pollution}, replace={Pollut.}]
			\step[fieldsource=\regexp{$MAPLOOP}, match={Quarterly}, replace={Quart.}]
			\step[fieldsource=\regexp{$MAPLOOP}, match={Polymer}, replace={Polym.}]
			\step[fieldsource=\regexp{$MAPLOOP}, match={Radiation}, replace={Radiat.}]
			\step[fieldsource=\regexp{$MAPLOOP}, match={Polytechnic}, replace={Polytech.}]
			\step[fieldsource=\regexp{$MAPLOOP}, match={Radiology}, replace={Radiol.}]
			\step[fieldsource=\regexp{$MAPLOOP}, match={Practice}, replace={Pract.}]
			\step[fieldsource=\regexp{$MAPLOOP}, match={Reactor}, replace={React.}]
			\step[fieldsource=\regexp{$MAPLOOP}, match={Precision}, replace={Precis.}]
			\step[fieldsource=\regexp{$MAPLOOP}, match={Receivers}, replace={Receiv.}]
			\step[fieldsource=\regexp{$MAPLOOP}, match={Principles}, replace={Princ.}]
			\step[fieldsource=\regexp{$MAPLOOP}, match={Recognition}, replace={Recognit.}]
			\step[fieldsource=\regexp{$MAPLOOP}, match={Proceedings}, replace={Proc.}]
			\step[fieldsource=\regexp{$MAPLOOP}, match={Record}, replace={Rec.}]
			\step[fieldsource=\regexp{$MAPLOOP}, match={Processing}, replace={Process.}]
			\step[fieldsource=\regexp{$MAPLOOP}, match={Rehabilitation}, replace={Rehabil.}]
			\step[fieldsource=\regexp{$MAPLOOP}, match={Conversion}, replace={Convers.}]
			\step[fieldsource=\regexp{$MAPLOOP}, match={Digital}, replace={Digit.}]
			\step[fieldsource=\regexp{$MAPLOOP}, match={Convention}, replace={Conv.}]
			\step[fieldsource=\regexp{$MAPLOOP}, match={Disclosure}, replace={Discl.}]
			\step[fieldsource=\regexp{$MAPLOOP}, match={Correspondence}, replace={Corresp.}]
			\step[fieldsource=\regexp{$MAPLOOP}, match={Discussions}, replace={Discuss.}]
			\step[fieldsource=\regexp{$MAPLOOP}, match={Critical}, replace={Crit.}]
			\step[fieldsource=\regexp{$MAPLOOP}, match={Dissertations}, replace={Diss.}]
			\step[fieldsource=\regexp{$MAPLOOP}, match={Crystal}, replace={Cryst.}]
			\step[fieldsource=\regexp{$MAPLOOP}, match={Distributed}, replace={Distrib.}]
			\step[fieldsource=\regexp{$MAPLOOP}, match={Crystallography}, replace={Crystallogr.}]
			\step[fieldsource=\regexp{$MAPLOOP}, match={Dynamics}, replace={Dyn.}]
			\step[fieldsource=\regexp{$MAPLOOP}, match={Cybernetics}, replace={Cybern.}]
			\step[fieldsource=\regexp{$MAPLOOP}, match={Earthquake}, replace={Earthq.}]
			\step[fieldsource=\regexp{$MAPLOOP}, match={Decision}, replace={Decis.}]
			\step[fieldsource=\regexp{$MAPLOOP}, match={Economics}, replace={Econ.}]
			\step[fieldsource=\regexp{$MAPLOOP}, match={Economic}, replace={Econ.}]
			\step[fieldsource=\regexp{$MAPLOOP}, match={Economical}, replace={Econ.}]
			\step[fieldsource=\regexp{$MAPLOOP}, match={Edition}, replace={Ed.}]
			\step[fieldsource=\regexp{$MAPLOOP}, match={Evolutionary}, replace={Evol.}]
			\step[fieldsource=\regexp{$MAPLOOP}, match={Education}, replace={Educ.}]
			\step[fieldsource=\regexp{$MAPLOOP}, match={Exhibition}, replace={Exhib.}]
			\step[fieldsource=\regexp{$MAPLOOP}, match={Electrical}, replace={Elect.}]
			\step[fieldsource=\regexp{$MAPLOOP}, match={Electric}, replace={Elect.}]
			\step[fieldsource=\regexp{$MAPLOOP}, match={Experimental}, replace={Exp.}]
			\step[fieldsource=\regexp{$MAPLOOP}, match={Electrification}, replace={Electrific.}]
			\step[fieldsource=\regexp{$MAPLOOP}, match={Exploratory}, replace={Explor.}]
			\step[fieldsource=\regexp{$MAPLOOP}, match={Electromagnetic}, replace={Electromagn.}]
			\step[fieldsource=\regexp{$MAPLOOP}, match={Exposition}, replace={Expo.}]
			\step[fieldsource=\regexp{$MAPLOOP}, match={Electroacoustic}, replace={Electroacoust.}]
			\step[fieldsource=\regexp{$MAPLOOP}, match={Express}, replace={Express}]
			\step[fieldsource=\regexp{$MAPLOOP}, match={Electronics}, replace={Electron.}]
			\step[fieldsource=\regexp{$MAPLOOP}, match={Electronic}, replace={Electron.}]
			\step[fieldsource=\regexp{$MAPLOOP}, match={Fabrication}, replace={Fabr.}]
			\step[fieldsource=\regexp{$MAPLOOP}, match={Emerging}, replace={Emerg.}]
			\step[fieldsource=\regexp{$MAPLOOP}, match={Faculty}, replace={Fac.}]
			\step[fieldsource=\regexp{$MAPLOOP}, match={Engineering}, replace={Eng.}]
			\step[fieldsource=\regexp{$MAPLOOP}, match={Engineers}, replace={Eng.}]
			\step[fieldsource=\regexp{$MAPLOOP}, match={Engineer}, replace={Eng.}]
			\step[fieldsource=\regexp{$MAPLOOP}, match={Ferroelectrics}, replace={Ferroelect.}]
			\step[fieldsource=\regexp{$MAPLOOP}, match={Environment}, replace={Environ.}]
			\step[fieldsource=\regexp{$MAPLOOP}, match={Francais, French}, replace={Fr.}]
			\step[fieldsource=\regexp{$MAPLOOP}, match={Equations}, replace={Equ.}]
			\step[fieldsource=\regexp{$MAPLOOP}, match={Frequency}, replace={Freq.}]
			\step[fieldsource=\regexp{$MAPLOOP}, match={Equipment}, replace={Equip.}]
			\step[fieldsource=\regexp{$MAPLOOP}, match={Foundation}, replace={Found.}]
			\step[fieldsource=\regexp{$MAPLOOP}, match={Ergonomics}, replace={Ergonom.}]
			\step[fieldsource=\regexp{$MAPLOOP}, match={Fundamental}, replace={Fundam.}]
			\step[fieldsource=\regexp{$MAPLOOP}, match={European}, replace={Eur.}]
			\step[fieldsource=\regexp{$MAPLOOP}, match={Generation}, replace={Gener.}]
			\step[fieldsource=\regexp{$MAPLOOP}, match={Evaluation}, replace={Eval.}]
			\step[fieldsource=\regexp{$MAPLOOP}, match={Geology}, replace={Geol.}]
			\step[fieldsource=\regexp{$MAPLOOP}, match={Geophysics}, replace={Geophys.}]
			\step[fieldsource=\regexp{$MAPLOOP}, match={Innovations}, replace={Innov.}]
			\step[fieldsource=\regexp{$MAPLOOP}, match={Innovation}, replace={Innov.}]
			\step[fieldsource=\regexp{$MAPLOOP}, match={Geoscience}, replace={Geosci.}]
			\step[fieldsource=\regexp{$MAPLOOP}, match={Institute}, replace={Inst.}]
			\step[fieldsource=\regexp{$MAPLOOP}, match={Institution}, replace={Inst.}]
			\step[fieldsource=\regexp{$MAPLOOP}, match={Graphics}, replace={Graph.}]
			\step[fieldsource=\regexp{$MAPLOOP}, match={Instrument}, replace={Instrum.}]
			\step[fieldsource=\regexp{$MAPLOOP}, match={Guidance}, replace={Guid.}]
			\step[fieldsource=\regexp{$MAPLOOP}, match={Instrumentation}, replace={Instrum.}]
			\step[fieldsource=\regexp{$MAPLOOP}, match={Harmonics}, replace={Harmon.}]
			\step[fieldsource=\regexp{$MAPLOOP}, match={Harmonic}, replace={Harmon.}]
			\step[fieldsource=\regexp{$MAPLOOP}, match={Insulation}, replace={Insul.}]
			\step[fieldsource=\regexp{$MAPLOOP}, match={History}, replace={Hist.}]
			\step[fieldsource=\regexp{$MAPLOOP}, match={Integrated}, replace={Integr.}]
			\step[fieldsource=\regexp{$MAPLOOP}, match={Horizon}, replace={Horiz.}]
			\step[fieldsource=\regexp{$MAPLOOP}, match={Intelligence}, replace={Intell.}]
			\step[fieldsource=\regexp{$MAPLOOP}, match={Hungary}, replace={Hung.}]
			\step[fieldsource=\regexp{$MAPLOOP}, match={Hungarian}, replace={Hung.}]
			\step[fieldsource=\regexp{$MAPLOOP}, match={Intelligent}, replace={Intell.}]
			\step[fieldsource=\regexp{$MAPLOOP}, match={Hydraulics}, replace={Hydraul.}]
			\step[fieldsource=\regexp{$MAPLOOP}, match={Interactions}, replace={Interact.}]
			\step[fieldsource=\regexp{$MAPLOOP}, match={Hydrology}, replace={Hydrol.}]
			\step[fieldsource=\regexp{$MAPLOOP}, match={Internationales}, replace={Int.}]
			\step[fieldsource=\regexp{$MAPLOOP}, match={International}, replace={Int.}]
			\step[fieldsource=\regexp{$MAPLOOP}, match={Illuminating}, replace={Illum.}]
			\step[fieldsource=\regexp{$MAPLOOP}, match={Isotopes}, replace={Isot.}]
			\step[fieldsource=\regexp{$MAPLOOP}, match={Imaging}, replace={Imag.}]
			\step[fieldsource=\regexp{$MAPLOOP}, match={Israel}, replace={Isr.}]
			\step[fieldsource=\regexp{$MAPLOOP}, match={Industrial}, replace={Ind.}]
			\step[fieldsource=\regexp{$MAPLOOP}, match={Japan}, replace={Jpn.}]
			\step[fieldsource=\regexp{$MAPLOOP}, match={Information}, replace={Inf.}]
			\step[fieldsource=\regexp{$MAPLOOP}, match={Journal}, replace={J.}]
			\step[fieldsource=\regexp{$MAPLOOP}, match={Informatics}, replace={Inform.}]
			\step[fieldsource=\regexp{$MAPLOOP}, match={Knowledge}, replace={Knowl.}]
			\step[fieldsource=\regexp{$MAPLOOP}, match={Laboratory}, replace={Lab.}]
			\step[fieldsource=\regexp{$MAPLOOP}, match={Laboratories}, replace={Lab.}]
			\step[fieldsource=\regexp{$MAPLOOP}, match={Mathematical}, replace={Math.}]
			\step[fieldsource=\regexp{$MAPLOOP}, match={Language}, replace={Lang.}]
			\step[fieldsource=\regexp{$MAPLOOP}, match={Mathematics}, replace={Math.}]
			\step[fieldsource=\regexp{$MAPLOOP}, match={Abstracts}, replace={Abstr.}]
			\step[fieldsource=\regexp{$MAPLOOP}, match={Analysis}, replace={Anal.}]
			\step[fieldsource=\regexp{$MAPLOOP}, match={Academy}, replace={Acad.}]
			\step[fieldsource=\regexp{$MAPLOOP}, match={Annals}, replace={Ann.}]
			\step[fieldsource=\regexp{$MAPLOOP}, match={Accelerator}, replace={Accel.}]
			\step[fieldsource=\regexp{$MAPLOOP}, match={Annual}, replace={Annu.}]
			\step[fieldsource=\regexp{$MAPLOOP}, match={Acoustics}, replace={Acoust.}]
			\step[fieldsource=\regexp{$MAPLOOP}, match={Apparatus}, replace={App.}]
			\step[fieldsource=\regexp{$MAPLOOP}, match={Active}, replace={Act.}]
			\step[fieldsource=\regexp{$MAPLOOP}, match={Applications}, replace={Appl.}]
			\step[fieldsource=\regexp{$MAPLOOP}, match={Administration}, replace={Admin.}]
			\step[fieldsource=\regexp{$MAPLOOP}, match={Applied}, replace={Appl.}]
			\step[fieldsource=\regexp{$MAPLOOP}, match={Administrative}, replace={Administ.}]
			\step[fieldsource=\regexp{$MAPLOOP}, match={Approximate}, replace={Approx.}]
			\step[fieldsource=\regexp{$MAPLOOP}, match={Advanced}, replace={Adv.}]
			\step[fieldsource=\regexp{$MAPLOOP}, match={Advances}, replace={Adv.}]
			\step[fieldsource=\regexp{$MAPLOOP}, match={Archives}, replace={Arch.}]
			\step[fieldsource=\regexp{$MAPLOOP}, match={Archive}, replace={Arch.}]
			\step[fieldsource=\regexp{$MAPLOOP}, match={Aeronautics}, replace={Aeronaut.}]
			\step[fieldsource=\regexp{$MAPLOOP}, match={Artificial}, replace={Artif.}]
			\step[fieldsource=\regexp{$MAPLOOP}, match={Aerospace}, replace={Aerosp.}]
			\step[fieldsource=\regexp{$MAPLOOP}, match={Assembly}, replace={Assem.}]
			\step[fieldsource=\regexp{$MAPLOOP}, match={Affective}, replace={Affect.}]
			\step[fieldsource=\regexp{$MAPLOOP}, match={Association}, replace={Assoc.}]
			\step[fieldsource=\regexp{$MAPLOOP}, match={Africa}, replace={Afr.}]
			\step[fieldsource=\regexp{$MAPLOOP}, match={African}, replace={Afr.}]
			\step[fieldsource=\regexp{$MAPLOOP}, match={Astronomy}, replace={Astron.}]
			\step[fieldsource=\regexp{$MAPLOOP}, match={Aircraft}, replace={Aircr.}]
			\step[fieldsource=\regexp{$MAPLOOP}, match={Astronautics}, replace={Astronaut.}]
			\step[fieldsource=\regexp{$MAPLOOP}, match={Algebraic}, replace={Algebr.}]
			\step[fieldsource=\regexp{$MAPLOOP}, match={Astrophysics}, replace={Astrophys.}]
			\step[fieldsource=\regexp{$MAPLOOP}, match={American}, replace={Amer.}]
			\step[fieldsource=\regexp{$MAPLOOP}, match={Atmosphere}, replace={Atmos.}]
			\step[fieldsource=\regexp{$MAPLOOP}, match={Atomic}, replace={At.}]
			\step[fieldsource=\regexp{$MAPLOOP}, match={Atoms}, replace={At.}]
			\step[fieldsource=\regexp{$MAPLOOP}, match={Broadcasting}, replace={Broadcast.}]
			\step[fieldsource=\regexp{$MAPLOOP}, match={Australasian}, replace={Australas.}]
			\step[fieldsource=\regexp{$MAPLOOP}, match={Bulletin}, replace={Bull.}]
			\step[fieldsource=\regexp{$MAPLOOP}, match={Australia}, replace={Aust.}]
			\step[fieldsource=\regexp{$MAPLOOP}, match={Bureau}, replace={Bur.}]
			\step[fieldsource=\regexp{$MAPLOOP}, match={{Automatic }}, replace={Autom.}]
			\step[fieldsource=\regexp{$MAPLOOP}, match={Business}, replace={Bus.}]
			\step[fieldsource=\regexp{$MAPLOOP}, match={Automation}, replace={Automat.}]
			\step[fieldsource=\regexp{$MAPLOOP}, match={Canadian}, replace={Can.}]
			\step[fieldsource=\regexp{$MAPLOOP}, match={Automotive}, replace={Automot.}]
			\step[fieldsource=\regexp{$MAPLOOP}, match={Automobiles}, replace={Automob.}]
			\step[fieldsource=\regexp{$MAPLOOP}, match={Automobile}, replace={Automob.}]
			\step[fieldsource=\regexp{$MAPLOOP}, match={Ceramic}, replace={Ceram.}]
			\step[fieldsource=\regexp{$MAPLOOP}, match={Chemical}, replace={Chem.}]
			\step[fieldsource=\regexp{$MAPLOOP}, match={Behavioral}, replace={Behav.}]
			\step[fieldsource=\regexp{$MAPLOOP}, match={Behavior}, replace={Behav.}]
			\step[fieldsource=\regexp{$MAPLOOP}, match={Chinese}, replace={Chin.}]
			\step[fieldsource=\regexp{$MAPLOOP}, match={Belgian}, replace={Belg.}]
			\step[fieldsource=\regexp{$MAPLOOP}, match={Climatology}, replace={Climatol.}]
			\step[fieldsource=\regexp{$MAPLOOP}, match={Biochemical}, replace={Biochem.}]
			\step[fieldsource=\regexp{$MAPLOOP}, match={Clinical}, replace={Clin.}]
			\step[fieldsource=\regexp{$MAPLOOP}, match={Bioinformatics}, replace={Bioinf.}]
			\step[fieldsource=\regexp{$MAPLOOP}, match={Cognitive}, replace={Cogn.}]
			\step[fieldsource=\regexp{$MAPLOOP}, match={Biology, Biological}, replace={Biol.}]
			\step[fieldsource=\regexp{$MAPLOOP}, match={Colloquium}, replace={Colloq.}]
			\step[fieldsource=\regexp{$MAPLOOP}, match={Kolloquium}, replace={Kolloq.}]
			\step[fieldsource=\regexp{$MAPLOOP}, match={Biomedical}, replace={Biomed.}]
			\step[fieldsource=\regexp{$MAPLOOP}, match={Communications}, replace={Commun.}]
			\step[fieldsource=\regexp{$MAPLOOP}, match={Communication}, replace={Commun.}]
			\step[fieldsource=\regexp{$MAPLOOP}, match={Biophysics}, replace={Biophys.}]
			\step[fieldsource=\regexp{$MAPLOOP}, match={Compatibility}, replace={Compat.}]
			\step[fieldsource=\regexp{$MAPLOOP}, match={British}, replace={Brit.}]
			\step[fieldsource=\regexp{$MAPLOOP}, match={Components}, replace={Compon.}]
			\step[fieldsource=\regexp{$MAPLOOP}, match={Component}, replace={Compon.}]
			\step[fieldsource=\regexp{$MAPLOOP}, match={Computational}, replace={Comput.}]
			\step[fieldsource=\regexp{$MAPLOOP}, match={Delivery}, replace={Del.}]
			\step[fieldsource=\regexp{$MAPLOOP}, match={Computers}, replace={Comput.}]
			\step[fieldsource=\regexp{$MAPLOOP}, match={Computer}, replace={Comput.}]
			\step[fieldsource=\regexp{$MAPLOOP}, match={Department}, replace={Dept.}]
			\step[fieldsource=\regexp{$MAPLOOP}, match={Computing}, replace={Comput.}]
			\step[fieldsource=\regexp{$MAPLOOP}, match={Design}, replace={Des.}]
			\step[fieldsource=\regexp{$MAPLOOP}, match={Condensed}, replace={Condens.}]
			\step[fieldsource=\regexp{$MAPLOOP}, match={Detector}, replace={Detect.}]
			\step[fieldsource=\regexp{$MAPLOOP}, match={Conferences}, replace={Conf.}]
			\step[fieldsource=\regexp{$MAPLOOP}, match={Conference}, replace={Conf.}]
			\step[fieldsource=\regexp{$MAPLOOP}, match={Development}, replace={Develop.}]
			\step[fieldsource=\regexp{$MAPLOOP}, match={Congress}, replace={Congr.}]
			\step[fieldsource=\regexp{$MAPLOOP}, match={Differential}, replace={Differ.}]
			\step[fieldsource=\regexp{$MAPLOOP}, match={Consumer}, replace={Consum.}]
			\step[fieldsource=\regexp{$MAPLOOP}, match={Digest}, replace={Dig.}] 
			\step[fieldsource=\regexp{$MAPLOOP}, match={{ of the }}, replace={{ }}]
			\step[fieldsource=\regexp{$MAPLOOP}, match={{ of }}, replace={{ }}]
			\step[fieldsource=\regexp{$MAPLOOP}, match={{Of }}, replace={{ }}]
			\step[fieldsource=\regexp{$MAPLOOP}, match={{ on }}, replace={{ }}]
			\step[fieldsource=\regexp{$MAPLOOP}, match={{ On }}, replace={{ }}]
			\step[fieldsource=\regexp{$MAPLOOP}, match={{On }}, replace={{ }}]
			\step[fieldsource=\regexp{$MAPLOOP}, match={{ in }}, replace={{ }}]
			\step[fieldsource=\regexp{$MAPLOOP}, match=\regexp{\\\x{26}}, replace={{and}}] 
			\step[fieldsource=\regexp{$MAPLOOP}, match={{, and }}, replace={{, }}]
			\step[fieldsource=\regexp{$MAPLOOP}, match={{, And }}, replace={{, }}]
			\step[fieldsource=\regexp{$MAPLOOP}, match={{ and }}, replace={{ }}]
			\step[fieldsource=\regexp{$MAPLOOP}, match={{ And }}, replace={{ }}]
			\step[fieldsource=\regexp{$MAPLOOP}, match={{ In }}, replace={{ }}]
			\step[fieldsource=\regexp{$MAPLOOP}, match={{In }}, replace={{ }}]
			\step[fieldsource=\regexp{$MAPLOOP}, match={{ the }}, replace={{ }}] 
			\step[fieldsource=\regexp{$MAPLOOP}, match={{ The }}, replace={{ }}] 
			\step[fieldsource=\regexp{$MAPLOOP}, match={{The }}, replace={}] 
			\step[fieldsource=\regexp{$MAPLOOP}, match={{ for }}, replace={{ }}] 
			\step[fieldsource=\regexp{$MAPLOOP}, match={First}, replace={1st}]
			\step[fieldsource=\regexp{$MAPLOOP}, match={Second}, replace={2nd}]
			\step[fieldsource=\regexp{$MAPLOOP}, match={Third}, replace={3rd}]
			\step[fieldsource=\regexp{$MAPLOOP}, match={Fourth}, replace={4th}]
			\step[fieldsource=\regexp{$MAPLOOP}, match={Fifth}, replace={5th}]
			\step[fieldsource=\regexp{$MAPLOOP}, match={Sixth}, replace={6th}]
			\step[fieldsource=\regexp{$MAPLOOP}, match={Seventh}, replace={7th}]
			\step[fieldsource=\regexp{$MAPLOOP}, match={Eighth}, replace={8th}]
			\step[fieldsource=\regexp{$MAPLOOP}, match={Ninth}, replace={9th}]
			\step[fieldsource=\regexp{$MAPLOOP}, match={Tenth}, replace={10th}]
			\step[fieldsource=\regexp{$MAPLOOP}, match={Eleventh}, replace={11th}]
			\step[fieldsource=\regexp{$MAPLOOP}, match={Twelfth}, replace={12th}]
			\step[fieldsource=\regexp{$MAPLOOP}, match={Thirteenth}, replace={13th}]
			\step[fieldsource=\regexp{$MAPLOOP}, match={Fourteenth}, replace={14th}]
			\step[fieldsource=\regexp{$MAPLOOP}, match={Fifteenth}, replace={15th}]
			\step[fieldsource=\regexp{$MAPLOOP}, match={Sixteenth}, replace={16th}]
			\step[fieldsource=\regexp{$MAPLOOP}, match={Seventeenth}, replace={17th}]
			\step[fieldsource=\regexp{$MAPLOOP}, match={Eighteenth}, replace={18th}]
			\step[fieldsource=\regexp{$MAPLOOP}, match={Nineteenth}, replace={19th}]
			\step[fieldsource=\regexp{$MAPLOOP}, match={Twentieth}, replace={20th}]
		}
	}
}

\usepackage{hologo}
\usepackage{listings}
\title{\LARGE \bf
	Showcasing Automated Vehicle Prototypes: A Collaborative Release Process to Manage and Communicate Risk*
}

\author{Marvin Loba$^{1\orcidlink{0000-0002-3425-5218}}$, Robert Graubohm$^{1\orcidlink{0000-0002-6682-4788}}$, and Markus Maurer$^{1\orcidlink{0000-0002-5357-9701}}$
		\thanks{*This research was accomplished within the project ``\unicaragil'' (FKZ~16EMO0285) and is continued within the project ``\autotech'' (FKZ~01IS22088R). We acknowledge the financial support for both projects by the Federal Ministry of Education and Research of Germany (BMBF).}
		\thanks{$^{1}$All authors are with the Institute of Control Engineering at Technische Universit\"at Braunschweig, 38106 Braunschweig, Germany
			{\tt\small \{loba,graubohm,maurer\}@ifr.ing.tu-bs.de}}%
	}
\begin{document}
	
\twocolumn[
\begin{@twocolumnfalse}
	\Huge {IEEE copyright notice} \\ \\
	\large {\copyright\ 2025 IEEE. Personal use of this material is permitted. Permission from IEEE must be obtained for all other uses, in any current or future media, including reprinting/republishing this material for advertising or promotional purposes, creating new collective works, for resale or redistribution to servers or lists, or reuse of any copyrighted component of this work in other works.} \\ \\
	{\Large Published in \emph{2024 IEEE 27th International Conference on Intelligent Transportation Systems (ITSC)}, Edmonton, Canada, September 24-27, 2024} \\ \\ 
	{\Large DOI: \href{https://doi.org/10.1109/ITSC58415.2024.10919548}{10.1109/ITSC58415.2024.10919548}} \\ \\ 
	
	Cite as: \vspace{0.2cm} \newline
		\vspace{0.2cm}
		\noindent\fbox{%
			\parbox{\textwidth}{%
				M.~Loba, R.~Graubohm, and M.~Maurer, ``Showcasing Automated Vehicle Prototypes: A Collaborative Release Process to Manage and Communicate Risk,'' in \emph{2024 IEEE Int. Conf. Intelligent Transportation Systems}, Edmonton, AB, Canada, 2024, pp. 3425-3432, doi: {10.1109/ITSC58415.2024.10919548}.
			}%
		}
		\vspace{2cm}
		
	\end{@twocolumnfalse}
	]
	
	\noindent\begin{minipage}{\textwidth}
		
		\hologo{BibTeX}:
		\footnotesize
		\begin{lstlisting}[frame=single]
			@inproceedings{loba_2024,
				author={{Loba}, Marvin and {Graubohm}, Robert and {Maurer}, Markus},
				title={Showcasing {Automated} {Vehicle} {Prototypes}: {A} {Collaborative} {Release} {Process} to {Manage} and {Communicate} {Risk}},
				address = {Edmonton, Canada},
				year={2024},
				pages = {325-3432},
				doi = {10.1109/ITSC58415.2024.10919548},
				publisher={IEEE}
			}
		\end{lstlisting}
	\end{minipage}

\maketitle
\thispagestyle{empty}
\pagestyle{empty}


\begin{abstract}
	The development and deployment of automated vehicles pose major challenges for manufacturers to this day. 
	Whilst central questions, like the issue of ensuring a sufficient level of safety, remain unanswered, prototypes are increasingly finding their way into public traffic in urban areas.
	Although safety concepts for prototypes are addressed in literature, published work hardly contains any dedicated considerations on a systematic release for their operation.
	In this paper, we propose an incremental release process for public demonstrations of prototypes' automated driving functionality.
	We explicate release process requirements, derive process design decisions, and define actors' tasks.
	Furthermore, we reflect on practical insights gained through implementing the release process as part of the \unicaragil\ research project, in which four prototypes based on novel vehicle concepts were built and demonstrated to the public.
	One observation is the improved quality of internal risk communication, achieved by dismantling information asymmetries between stakeholders.
	Design conflicts are disclosed -- providing a contribution to nurture transparency and, thereby, supporting a valid basis for release decisions.
	We argue that our release process meets two important requirements, as the results suggest its applicability to the domain of automated driving and its scalability to different vehicle concepts and organizational structures.
	
\end{abstract}


\section{INTRODUCTION}
\label{sec:intro}

The widespread introduction of series vehicles equipped with automated driving systems is still pending. 
Yet, prototypes are gradually finding their way from proving grounds into public traffic.
The operation of automated vehicles in the open-world context of urban traffic is always subject to an \emph{inherent risk} that stems from functional and systemic causes, e.g., technical limitations and incomplete requirements. 
This inherent risk can be reduced but never eliminated~\cite{maurer2018}.
We claim that the unavoidable existence of residual risk also applies to prototypes, which by their very nature are innovative complex systems in which safety is an emergent property.

Unfortunately, prototypes were involved in multiple accidents in previous years, e.g., involving the companies Uber, Pony.AI, or Cruise~\cite[III.B.]{wansley2024}.
Recall investigations after a major incident caused by a prototype from Cruise even led to the Department of Motor Vehicles in California suspending the permit for driverless test operation in October 2023~\cite{DMV2023}.
The (social) media responses following such incidents indicate the need for an open discussion on the level of risk, posed by prototypes' operation, that is acceptable for society.

Concerning vehicles providing automatic emergency braking, \textcite{homann2002} demanded in 2002 already that an open discussion on risk is held with stakeholders in society before systems are launched on the market~(translated from the reference to \cite{homann2002} by \textcite{maurer2021}).
In 2016, \textcite{wachenfeld2016} stated that with the first accident caused by an automated vehicle its release will be questioned, emphasizing that the basis for a release should be designed transparently.

Challenging the basis for a release resonates with the question of what ``safety'' actually means.
One definition common in the field is the ``absence of unreasonable risk''~\cite[Part 1, 3.132]{ISO_2018} but \textcite{salemTobepublished} underpin deviating views stakeholders have on ``safety'' and ``risk.''
Regarding conceptual uncertainty, \textcite{fleischer2023} argues linguistically.
Accordingly, ``safety'' is a common language concept, usually associated with an intuitive interpretation for each stakeholder and consensual in the expectation that automated vehicles must be ``safe.''
Implicit understandings may, however, only lead to a superficial consensus on the meaning individuals attribute to ``safety,'' suggesting that stakeholders also have divergent understandings of ``safe'' prototype operation.
A need arising from this idea is to strengthen the communication between various domain experts involved in developing and deploying prototypes.
We claim that a stakeholder-collaborative release process, which guides development, can account for this.

In this paper, with ``release'' we refer to the granting of permission for a specific prototype operation by decision makers within an organization. 
This does not include certification/type approval by regulatory authorities, whose involvement is also mandatory in certain countries to obtain a permit for prototypes' operation on public roads -- cf. for instance \cite[§ 16]{AFGBV2022} for testing permit requirements in Germany.
These cannot assumed to be guaranteed (e.g., regarding hardware integrity) by novel early-stage prototypes to be released for demonstration purposes as part of research projects, which reflects the scope of this paper.

With this paper, we seek to accomplish two goals.
First, we aim to stimulate a discussion on the level of safety to be achieved during development.
With respect to series vehicles, the debate on defining safety is already underway, e.g., as ``hot potato'' in the focus field ``safety and risk'' of the German Round Table for Autonomous Driving~\cite{maurer2023}.
Considering prototypes, we perceive that the debate is currently missing. 
But we believe that a consensus on the level of reasonable residual risk is a prerequisite for responsible authorities within an organization to be able to make conscientious decisions as to whether a prototype can be released to enter its intended operation, e.g., for demonstration purposes.

\begin{figure}[!h]
	\centering
	\includegraphics[scale=0.1]{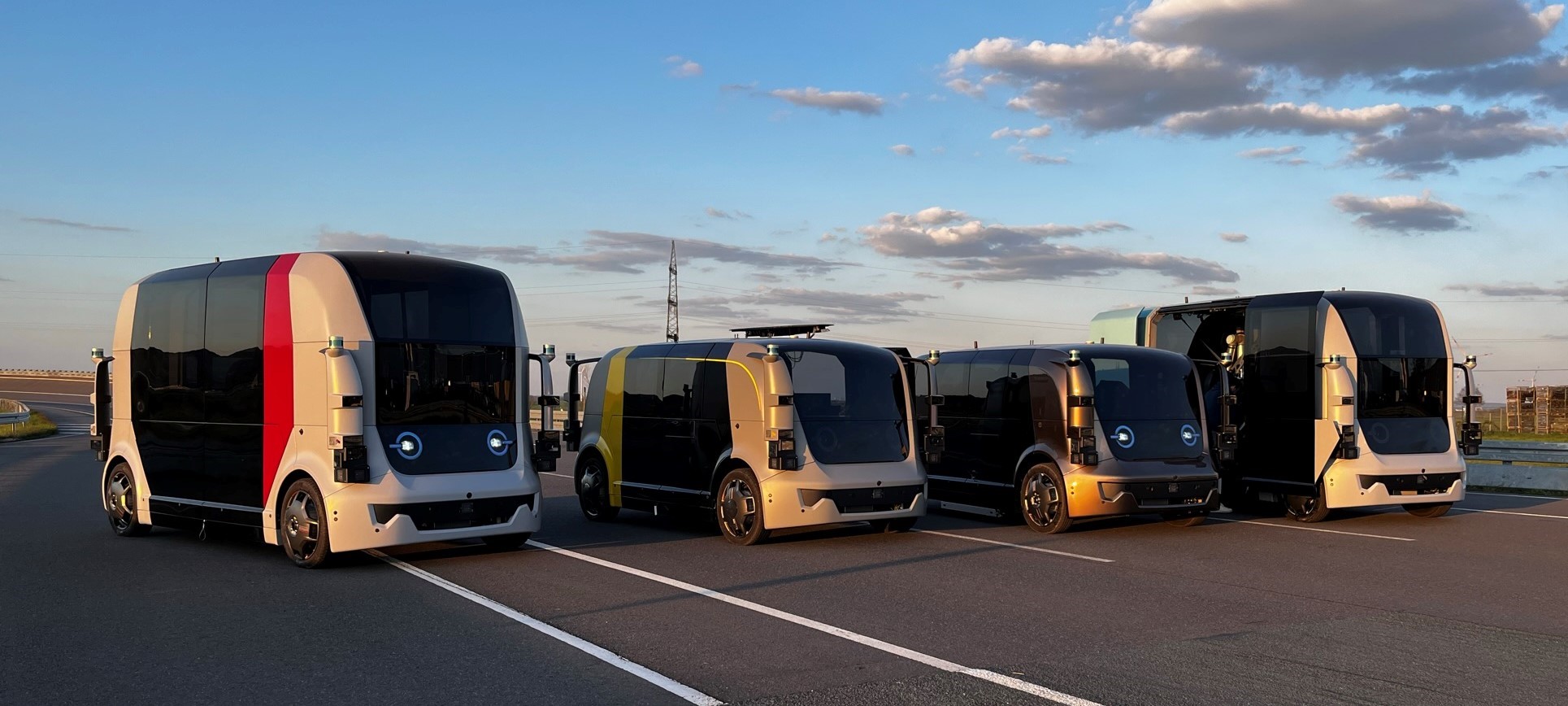}
	\caption{Prototypes built and demonstrated in the project \unicaragil; left to right: \autoshuttle, \autotaxi, \autoelf, and \autocargo\ \cite{vankempen2023}.}
	\label{fig:UNICAR}
\end{figure}

Second, we aim to tackle the scarcity of published knowledge on a systematic release of prototypes and provide entities with means to prepare the basis for a release.
This includes the disclosure of the risk reduction truly achieved by implementation.
To this end, we propose a release process that we designed and conducted as part of the research project \unicaragil.
Innovative vehicle concepts were examined in the project \cite{vankempen2023}. 
Four prototypes  (\cref{fig:UNICAR}) representing different use cases were developed from scratch and built by a large consortium, with minimal recourse to legacy knowledge and without a fully developed and safety-assured base vehicle.
Hence, the prototypes relied on novel components that lack series integrity.
As a result, a systematic release process played a key role to manage and communicate risk.
The prototypes were demonstrated in driverless operation on a test track to the public in May 2023, with passengers in three prototypes.

It is important to clarify that this paper's focus is not on the definition of appropriate technical and organizational measures to reduce the risk for the prototypes' demonstration.
Although the design and realization of the safety concept was driven by the release process and, thus, is also covered in this paper, we do not address the project-specific safety concept in depth. 
Rather, the safety concept represents one of multiple artifacts that form the basis of a release within the presented process framework, as detailed in \cref{subsec:ProcessDesign_GeneratingEvidence}.

In the remainder of this paper, first, we cover related work (\cref{sec:RW_ReleaseProcess}). 
Then, we present the release process design and realization in detail~(\cref{sec:DesignRealization}). 
Finally, we reflect on the release process execution, discussing the experience we gained (\cref{sec:Discussion}) before concluding our paper (\cref{sec:Conclusion}).


\section{RELATED WORK}
\label{sec:RW_ReleaseProcess}

Since insights on manufacturers' processes underlying series vehicle releases are internal to the organizations and not openly accessible, it is not feasible to orient our release process to series development procedures.

Regarding the release of prototypes a distinction is helpful, as they can exhibit different levels of maturity:
On the one hand, more mature prototypes exist that operate at high frequencies in less restricted operating environments and may be considered rather as pre-series vehicles.
For instance, prototypes from the company Waymo operate in fleets on public roads in selected North American cities. 
Waymo explains that ``each change of software undergoes a rigorous release process'' including simulation, closed-course tests, and driving on public roads~\cite{waymollc2021}.
However, specific requirements/procedures for moving from testing facilities to on-road testing or omitting safety drivers are not disclosed -- and especially release documents are not published.

On the other hand, prototypes in research contexts have been demonstrated in controlled environments for at least 40 years~\cite{dickmanns1987,thorpe1988}.
Literature on such demonstrations mainly deals with technical/organizational measures, i.e., a safety concept, where human controllability acts as central risk mitigation mechanism.
Controllability is either supported by interfaces providing humans with the capability to overrule actuators~\cite{thorpe1998,ziegler2014,nothdurft2011} or to intervene by using remote stop systems~\cite{broggi2013,jan2021}.
However, these references merely allow us to assume that an assessment of the safety concept by a decision maker (whether according implementation results in sufficient risk reduction) served as basis for approving the prototypes' operation.
As presented by \textcite{bagschik2018,nothdurft2011}, this assessment may be supported by an external review of a certification agency.
While a reliable safety concept is a key factor for weighing a release decision, the aforementioned references are barely applicable to our work since the publications do not propose a systematic release process for prototypes.

To the best of our knowledge, the only source actually closely related to our work comes from \textcite{strauss2023}.
The authors use the example of a driverless shuttle to illustrate a release process for a specific operational context. 
Accordingly, the decision of a release authority is based on, among other things, extensive documentation of tests and safety measures.
Yet, the authors do not address release process actors and their interaction in detail, as they are focusing on a high-level release process chronology, the analysis of the intended operation environment, safety analyses, and details on the technical realization of the driverless shuttle prototype.
We strongly encourage to get in touch with us if there is any further relevant literature that is missing from our review and helps to resolve the issues outlined in this paper.


\section{DESIGN AND REALIZATION OF A RELEASE PROCESS FOR PROTOTYPE DEMONSTRATION}
\label{sec:DesignRealization}

In this section, we propose the release process designed and implemented in the \unicaragil\ project. 
To this end, we address requirements and associated release process design decisions (\cref{subsec:ProcessDesign_Reqs}). 
Then, we cover involved actors and the process workflow that results from the design decisions taken (\cref{subsec:ProcessDesign_Workflow}). 
Finally, in \cref{subsec:ProcessDesign_GeneratingEvidence}, we explain the creation of release modules that provide evidence for systematic risk reduction and, compiled to release documents, serve as a basis for the release for public demonstration.

\subsection{Requirements and Derived Process Design Decisions}
\label{subsec:ProcessDesign_Reqs}

\noindent For prototypes in a research project context, we consider
\begin{itemize}
	\item knowledge asymmetry between different stakeholders,
	\item parallelism of \mbox{top-down} safety analyses and \mbox{bottom-up} function and component development,
	\item no developed and safety-assured base vehicle, and
	\item lack of series integrity of novel prototypical components
\end{itemize}
as major challenges for both prototype development and the establishment of a release process for their demonstration.

With respect to development processes, one widely known model for developing mechatronic systems is the \mbox{V-model}.
While ``classic'' V-model schemes do not depict an iterative process character, the guideline VDI 2206 points out that several macrocycle runs can be required to achieve the final product~\cite{vdi2006}. 
Accordingly, prototypes can represent one kind of intermediate product that results from completing the first development cycles.
However, the \mbox{V-model} illustrates a development context for which system-wide requirements are known at the outset and no feedback loops are required~\cite{zurawka2016}.
Therefore, we consider following a sequential V-model unsuitable to guide the development of innovative and complex systems.
In contrast, the requirement definition for automated vehicle prototypes should be an evolutionary refinement.
Hence, the release process shall allow for an agile development approach that promotes iterations in early phases, in which prototypes can be allocated.

The release process design is based on an expert-based requirement elicitation, enriched by experience of all project partners they gained from demonstrations that were successfully carried out in the past.
Captured requirements (\ReqBox) and related process implications are elaborated as follows:

\RequirementLine{Structured \& gradual}
To enable a structured process, a process chronology with distinct steps is determined.
We define a process workflow in advance (cf. \cref{subsec:ProcessDesign_Workflow}). 
To account for the novelty and complexity of the prototypes, we pursue a procedure that restricts the extension of the functional scope tested in operation to small steps. 
Hence, we foresee a gradual release, i.e., a plan for successive release stages oriented to the specified integration plan of the prototypes. 
Each stage (see \cref{tab:ReleaseStage}) is linked to conditions under which the prototypes are allowed to operate after a release is granted.

\newcolumntype{L}[1]{m{\dimexpr#1-2\tabcolsep-\arrayrulewidth}}
\begin{table}[!h]
	\renewcommand{\arraystretch}{1.3}
	\centering
	\caption{Definition of incremental release stages.}
	\label{tab:ReleaseStage}
	\begin{NiceTabular}{L{0.35\columnwidth}L{0.65\columnwidth}}[rules/color=rotLogo,rules/width=1pt]
		\CodeBefore
		\rowlistcolors{2}{white,graux}
		\Body		
		\hline
		\textbf{Release stage \newline \emph{Operating mode}$^{*}$} & \textbf{Detailed description of release stage}\\
		\hline  
		Stage 1 \newline \emph{Manual Operation} & Manual controlled rides on test sites with speeds of up to approximately \qty{5}{\kilo\meter\per\hour} \\
		Stage 2 \newline \emph{Manual Operation} &  Manual controlled rides on test sites \\
		Stage 3 \newline \emph{Automated Operation} & Testing of (automated driving) functions that require safety drivers as a fallback level in a controlled environment \\
		Stage 4 \newline \emph{Automated Operation} & Testing the demonstration without access for guests \\
		Stage 5 \newline \emph{Automated Operation} & Public demonstration on a test track \\ 
		\Hline
		\multicolumn{2}{@{}p{\columnwidth}@{}}{$^{*}$The operating modes ``Manual Operation'' and ``Automated Operation'' are discussed with respect to the project context by \textcite{jatzkowski2021}.} \\
	\end{NiceTabular}
\end{table}
\RequirementLine{Documented}
We find that the process must document both risk and risk mitigation measures comprehensibly. 
To obtain a reliable basis for release approval, i.e., the assessment of reasonable residual risk by an appointed committee, we introduce the concept of profound ``release documents.''
Yet, to ensure safety, the prototype release approval is based both on appropriate documentation within the release documents and on the appraisal of actual ``readiness'' of the prototypes. 

\begin{figure*}[!h]
	\centering
	\includegraphics[width=0.8\textwidth]{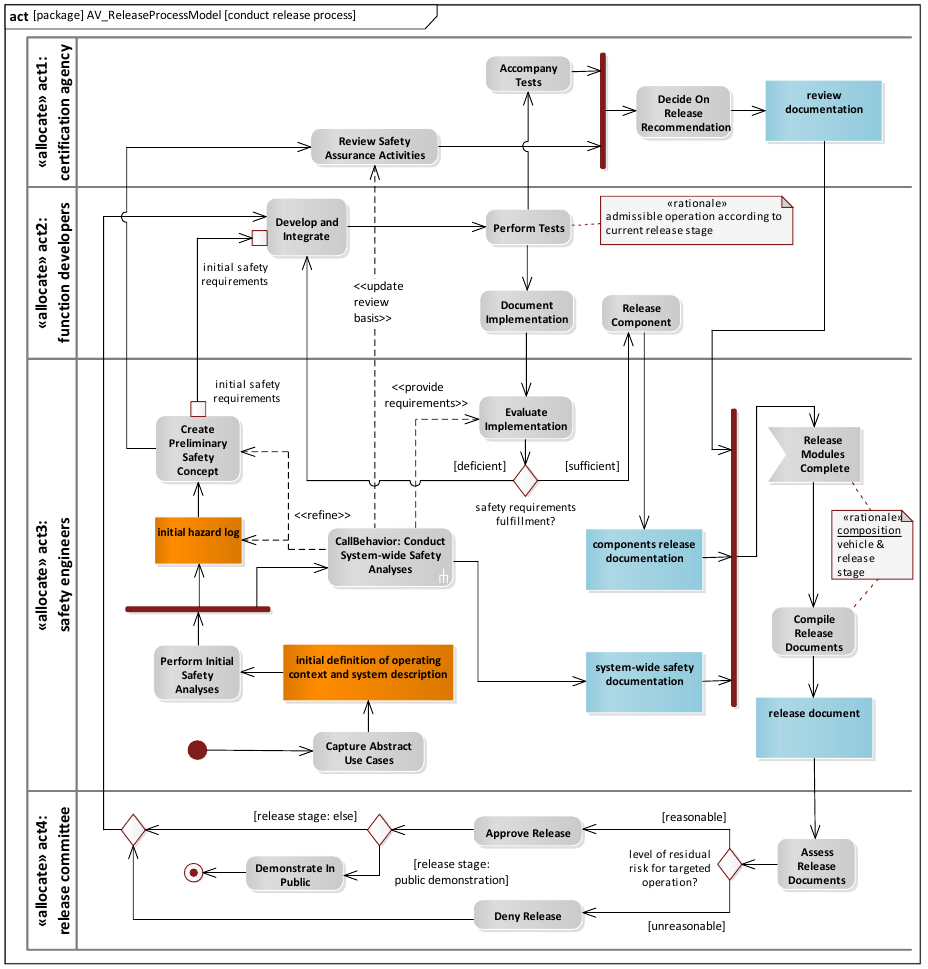}
	\caption[]{UML/SysML activity diagram that represents the activity ``conduct release process'' as a prescriptive illustration of the release process. \OrangeBox\ (SysML object) indicate initial work products, \GreyBox\ (SysML action) tasks, and \BlueBox\ (SysML object) documents.}
	\label{fig:ProcessFlow}
\end{figure*}

\RequirementLine{Measurable}
Guaranteeing a coordinated process execution that targets systematic risk reduction constitutes another process requirement.
All project participants must understand and bring about the prerequisites for ``safe'' operation. 
Maintaining a binding nature of the process by adhering to initial agreements is a decisive factor in achieving this goal.
Thus, we attune compositions of ``release modules'' for each release stage in the concept phase of prototype development already. 
These modules represent documented development evidence and form the building blocks of conclusive release documents. 
A variety of system-wide and component-level release modules are designated in the compositions (cf. \cref{subsec:ProcessDesign_GeneratingEvidence}). 
With a predefined composition of modules, we aim to foster measurability, as lack of progress can be traced back to explicit root issues hindering release for the next stage.

\RequirementLine{Accountability-driven}
Accountability is considered highly relevant to promote diligence and clarify on responsibility and, thereby, contribute to adherence to the schedule.
Accordingly, roles and associated tasks need to be defined unambiguously. 
We encourage accountability in the process by assigning specific design and test documentation obligations on component level to individual function developers. 
Beyond that, function developers need to actively release the components assigned to them. 
These documented component releases are included as modules in release documents for the aspired release stage.

\RequirementLine{Compliant \& harmonized}
We strive for a process that reflects the state of the art. 
On the one hand, this means that the release process shall follow a thorough \emph{Safety-by-design} paradigm.
A \mbox{Safety-by-Design} paradigm reflects a strong focus on dealing with safety requirements, involving the conduction of hazard analyses and specification of technical solutions that meet defined safety goals at an early stage. 
This contrasts with the assumption that extensive testing after product development would be sufficient to guarantee safety.
Hence, a focus lies on harmonizing ongoing function development on component level, which is partially rooted in early assumptions that are recorded in initial work products ({\raisebox{0mm}{\footnotesize\OrangeBox}} in \cref{fig:ProcessFlow}), and refined safety requirements on system level, which evolve from continuous system-wide safety analyses.

On the other hand, this refers to the application of standards in the field, e.g., vehicle-wide hazard analyses are guided by processes of the safety standards ISO~26262~\cite{ISO_2018} and ISO~21448~\cite{ISO_2022}.
External supervision supports compliance and increases confidence in the safety assurance activities.
Consequently, the process involves an independent certification agency with competence for respective audits evaluating the compliance with these standards.

\RequirementLine{Transparent}
Another requirement is cultivating transparency. 
Not only does the requirement arise that the release process is understood thoroughly by all release process actors but we also formulate explicit requirements directed towards the release modules:
The contained documentation must disclose the risk for operation, associated with uncertainties and limitations of the components as well as any emergent effects that may occur and cause hazardous behavior at the system level. 
Furthermore, risk reduction measures, safety strategies, and fallback mechanisms at the vehicle and component level need to be recorded within the release modules' documentation.
As we deem accessibility of the compiled release documents of central relevance to enable conscientious release decisions, the deliberate tailoring of language and structure of each release document is performed before an assessment of the documents is carried out.

\subsection{Overall Release Process Workflow and Involved Actors}
\label{subsec:ProcessDesign_Workflow}

\cref{fig:ProcessFlow} depicts the resulting release process designed and implemented in the project \unicaragil. 
Tasks and documents are allocated to four defined actors via swimlanes.\footnote{The corresponding actors \Actor\ allocated to the ``activity partitions'' are also represented in the block definition diagram in \cref{fig:ReleaseComposition}.} 
Accordingly, safety engineers perform initial safety analyses based on a project-wide agreed starting point, e.g., an abstract capturing of use cases and the envisioned operating environment. 
As part of these analyses, system-wide hazards identified early are recorded in an initial hazard log.
In this paper, according to \textcite{salem2024}, a hazard log is understood as an artifact that lists identified hazards and their mitigation status. 
A preliminary safety concept, which responds to these hazards, provides a first set of safety requirements as input to the function developers’ activities. 

The development based thereon is followed by \mbox{(sub-)system} integration and test case execution. 
As indicated by a ``rationale'' element, the currently approved release stage determines permissible test operation conditions. 
After testing, function developers describe the implementation and submit the related documentation to the safety engineers.

Parallel to the function developers’ activities, vehicle-wide safety analyses are continued by the safety engineers. 
These analyses allow to concretize the safety concept and associated work products continuously, resulting in an evolving safety documentation.
The generation of underlying release modules is detailed in \cref{subsec:ProcessDesign_GeneratingEvidence}. 
The refined safety concept acts as a basis for evaluating the ongoing implementation. 
Safety engineers examine the implementation documentation at component level that is provided by the function developers.
They check if sufficient proof is furnished that defined safety requirements served as basis of implementation and that those requirements are verified by testing. 
If a mismatch, i.e., deficient requirement fulfillment, is revealed, function developers are mandated to review their documentation and implementation.
Otherwise, function developers issue a component release, contributing to the aggregated components release documentation.

The certification agency collaborates with both the function developers (\emph{``Accompany Tests''}) and safety engineers (\emph{``Review Safety Assurance Activities''}).
The agency's recommendation for or against a release is captured in a review documentation that is also included in the release documents.

When a release document is compiled, it can be reviewed by a release committee. 
Depending on an assessment of the residual risk, this committee decides whether a release can be issued for the targeted release stage and the associated operating conditions. 
Approval of the final release stage ends the process by permitting the demonstration. 
Previous release stages address the use of the prototypes throughout advancing prototype construction and testing of their automated driving functionality, as illustrated in \cref{tab:ReleaseStage}.

\subsection{Evidence for the Prototypes' Safety Assurance}
\label{subsec:ProcessDesign_GeneratingEvidence}

\begin{figure*}[!h]
	\centering
	\includegraphics[width=.87\textwidth]{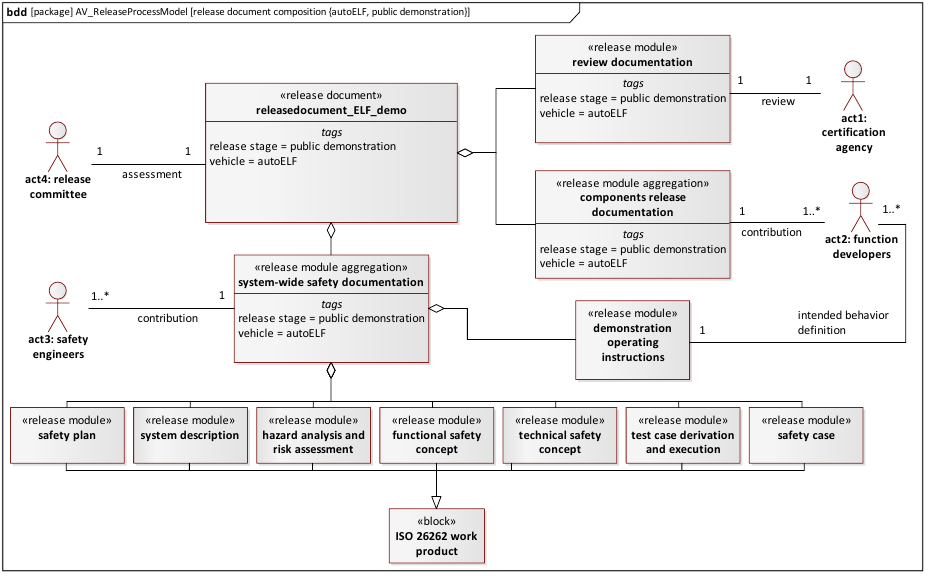}
	\vspace{-1em}
	\caption{SysML block definition diagram representing an exemplary release document composition (release stage: public demonstration, prototype: \autoelf).}
	\label{fig:ReleaseComposition}
	\vspace{-1em}
\end{figure*}

\cref{fig:ReleaseComposition} shows the schematic release document composition for public demonstration and one of the \unicaragil\ prototypes named \autoelf. 
There are two main aggregation points for a multitude of release modules, with the relationship modeled by logical aggregation (\AggregationRelShip). 
The first is the system-wide safety documentation and the second is the components release documentation, contributed by safety engineers and function developers, respectively.

\SecLine{System-wide safety documentation}
As indicated by the generalization relationship (\GeneralizationRelShip), except for the operating instructions, all release modules contributing to the system-wide documentation can be associated with matching ISO 26262 work products (highlighted in italics in this paragraph):
A \emph{safety plan} (cf.~\cite[Part 2, 6.4.6]{ISO_2018}) functions as coordination instrument for the safety assurance activities.
A system description (\emph{item definition}, cf.~\cite[Part 3, 5]{ISO_2018}) allows to perform the \emph{hazard analysis and risk assessment} (cf.~\cite[Part 3, 6]{ISO_2018}).
Functional safety requirements are derived from resulting safety goals and captured in the \emph{functional safety concept} (cf.~\cite[Part 3, 7]{ISO_2018}).
Technical safety requirements can be elicited and allocated to components that need to fulfill these requirements. 
The \emph{technical safety concept} (cf.~\cite[Part 4, 6]{ISO_2018}) comprises technical safety requirements and enables to deduce test cases (cf.~\cite[Part 4, 7-8]{ISO_2018}). 
Based on all of the work products, a \emph{safety case} (cf.~\cite[Part 2, 6.4.8]{ISO_2018}) is prepared, with the absence of unreasonable risk being the top-level claim, backed by a semi-formal argument about how the gathered development evidence supports this claim.

When it comes to identifying hazards, an approach to conduct the initial hazard identification was developed in the project context by \textcite{graubohm2020b}.
Considering vehicle dynamic and boarding scenarios representative for the intended operational scope was decisive in order to identify hazards. 
Coupling these representative operational scenarios with possible component malfunctions enabled us to derive hazardous scenarios and assess the associated risk.

\newcolumntype{C}[1]{>{\centering\arraybackslash}m{\dimexpr#1-2\tabcolsep-\arrayrulewidth}}
\newcolumntype{N}[1]{m{\dimexpr#1-2\tabcolsep-\arrayrulewidth}}
\begin{table}[!h]
	\renewcommand{\arraystretch}{1.6}
	\centering
	\caption{Definition of Research Safety Integrity Levels.}
	\label{tab:RSIL}
	\begin{NiceTabular}{C{0.11\columnwidth}N{0.72\columnwidth}|C{0.17\columnwidth}}[rules/color=rotLogo,rules/width=1pt]
		\CodeBefore
		\rowlistcolors{2}{white,graux}
		\Body		
		\hline
		\textbf{RSIL} & \textbf{Definition of the respective RSIL} &\textbf{ASIL\newline Reference$^{*}$}\\
		\hline  
		RSIL 0 & A conscientious execution of safety assurance activities is considered sufficient to control the risk. There is no need for additional measures. & QM \\
		RSIL 1 & Risk is classified as \textbf{\emph{\textcolor{rotLogo}{very low}}}. Appropriate technical or organizational measures must be defined, implemented, and tested to address the risk.  & ASIL A\\
		RSIL 2 & Risk is classified as \textbf{\emph{\textcolor{rotLogo}{low}}}. Appropriate technical or organizational measures must be defined, implemented, and tested to address the risk. & ASIL B\\
		RSIL 3 & Risk is classified as \textbf{\emph{\textcolor{rotLogo}{high}}}. Appropriate technical or organizational measures must be defined, implemented, and tested to address the risk. & ASIL C\\
		RSIL 4 &  Risk is classified as \textbf{\emph{\textcolor{rotLogo}{very high}}}. Appropriate technical or organizational measures must be defined, implemented, and tested to address the risk. & ASIL D\\
		\Hline
		\multicolumn{3}{@{}p{\columnwidth}@{}}{$^{*}$ASIL is assessed based on the risk parameters \emph{exposure}, \emph{controllability}, and \emph{severity} (cf. \cite[6.4.3]{ISO_2018}) and converted to RSIL as defined.}
	\end{NiceTabular}
\end{table} 

In order to assess this risk, we introduced the concept of RSIL (\textbf{R}esearch \textbf{S}afety \textbf{I}ntegrity \textbf{L}evel, see \cref{tab:RSIL}), targeting a qualitative delimitation of levels that constitute the risk potential of hazardous scenarios.
This classification aims to provide an indicator of the ``priority'' with which hazardous scenarios, respectively safety requirements derived from them, need to be addressed by development effort during implementation.
The underlying assumption is that conscientious development according to the categorization results in an overall risk reduction to a reasonable threshold.
As series integrity cannot be assumed for the prototype realization in the project, no list of prescribed methods for dealing with systematic faults, e.g., with regard to testing hardware and software, or random hardware faults, e.g., permissible failure rates, are linked to the RSILs, in contrast to ASIL (\textbf{A}utomotive \textbf{S}afety \textbf{I}ntegrity \textbf{L}evel \cite[Part 1,~3.6]{ISO_2018}) recommendations provided by the ISO~26262 standard.

Regarding the safety concept, \textcite{stolte2020} discuss the linking of identified hazards and emerging safety goals with project-specific safety mechanisms in detail.
It is worth mentioning that prototypical safety functions that were investigated in the project, such as the function ``safe halt,'' (cf. \cite{ackermann2023}) implemented for demonstration purposes, could not contribute to the safety concept. 
On the contrary, these functions had to be considered as causes of malfunctioning behavior, possibly leading to additional safety requirements.

From a technical perspective, the safety concept in the project relied heavily on redundancies and fallback strategies in case of undesirable component behavior. 
Defining organizational measures to deal with safety requirements is highly important for prototypical on-road testing and demonstrating in a controlled environment, as it partially accounts for the lack of component integrity.
Examples for organizational measures correspond to controlling the operational design domain, for instance via the definition of possible encounter traffic or preventing access of external persons to the driving corridors of the prototypes.
Additionally, one instrument we found to be essential for the design of a resilient safety concept in the project was a ``safety watch.'' 
A radio emergency stop system allowed human track marshals to stop a prototype immediately.
Besides functional requirements, non-functional aspects like mechanical safety or electromagnetic compatibility were also addressed in the safety concept. 
Relevant test cases for the verification and validation of safety requirements involved simulation as well as executing fault injection and demonstration scenario tests.

Complementary to the safety concept, operating instructions for testing and demonstration were prepared, containing, e.g., detailed descriptions of (incident) procedures and roles. 
With the operating instructions, we aimed to \mbox{reasonably} calibrate the trade-off between demonstration scope, e.g., complexity of the demonstrated functionality, and risk for operation. 
The instructions included the routes for the prototypes’ demonstrations, which were planned based on an assessment of the prototypes' capabilities and limitations.

\begin{figure}[!h]
	\centering
	\vspace{-.5em}
	\includegraphics[scale=0.55]{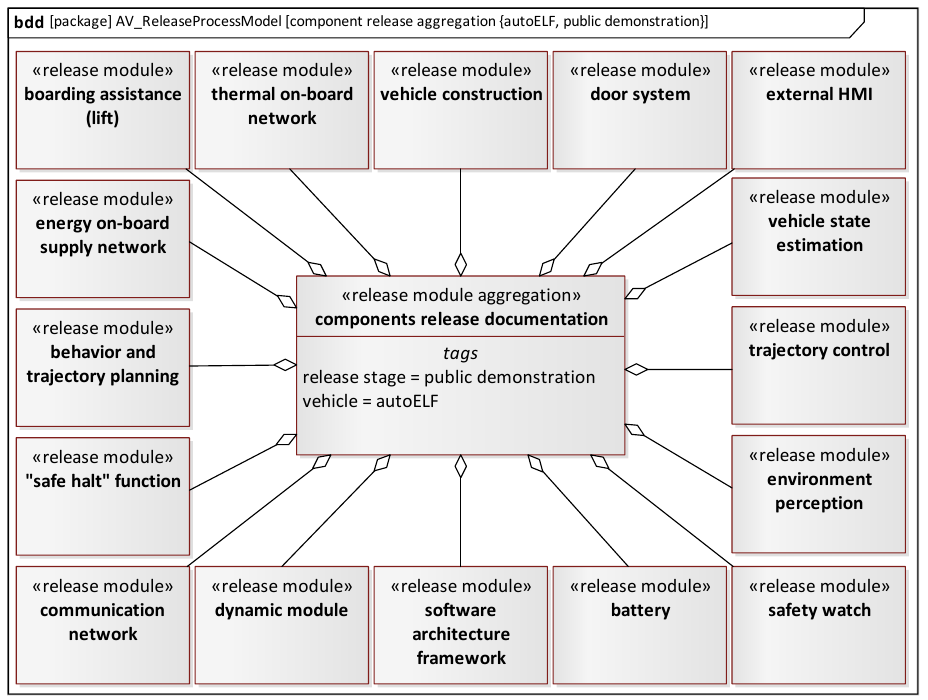}
	\caption{SysML block definition diagram representing the composition of an example component release documentation by various release modules (release stage: public demonstration, prototype: \autoelf).}
	\label{fig:componentreleasesL5}
	\vspace{-1em}
\end{figure}

\SecLine{Components release documentation}
On the function developers' side, various release modules had to be contributed at the component level. 
As an example, for release stages that require automated operation developers had to document component releases for environment perception, behavior planning, and motion control. 
These and other modules shown in \cref{fig:componentreleasesL5} represent functional architecture elements common in a sense-plan-act control scheme and, thus, were deemed necessary for releasing prototypes and using them in automated operation mode.

Besides component releases for shared components, prototype-specific components and functions led to differentiated compositions for the respective release documents.
For instance, the \autoelf\ was designed to provide the use case of an autonomous family vehicle.
To achieve the necessary accessibility for impaired and/or older family members, the prototype \autoelf\ was equipped with (an actuated) lift platform that allowed for boarding the \autoelf\ barrier-free.
Hence, the release module ``boarding assistance (lift)'' was required from the developers to compile the components release documentation for the prototype \autoelf\ -- as can be seen in~\cref{fig:componentreleasesL5} for the release stage \textit{public demonstration}.

Correspondingly, the composition of release modules primarily resulted from architectural considerations with respect to the functions required for the prototype operation that was foreseen for the release stage.
Component release documents were based on a project-wide template to nurture consistency and coherence of the function developers' documentation. 
Accordingly, function developers had to describe the component's functions including their interfaces and subsystem boundaries, implemented fallback mechanisms, and known limitations. 
Also, developers had to clarify on hazards caused by the component, safety-relevant component requirements, and derived mitigation strategies.
Furthermore, component level tests had to be recorded. 
Traceability between \mbox{top-down} and \mbox{bottom-up} considerations was fostered, as the technical safety concept explicitly linked technical safety requirements to the corresponding component release documentation.

\section{DISCUSSION}
\label{sec:Discussion}

Operationalizing the release process in a specific project led to valuable insights. 
We observed that assigning documentation obligations promoted accountability and timely contributions. 
While the prototypes were largely based on the same architecture and shared platforms, they also featured individual functions.
The presented process takes these differences into account, in particular via the modularity provided by prototype-specific compositions of required component releases. 
The successful process execution suggests not only the general applicability of the process, but also the potential of scalability for deviating vehicle concepts.

Due to the alignment with processes of established standards such as ISO~26262, we consider the process to be also applicable to conventional vehicle prototypes.
Shortcomings with regard to the process design that can be addressed in the future (e.g., to cover testing permit requirements for operation on public roads) include security considerations, accounting for design conflicts, or alternative means to represent risk potential of hazardous scenarios than RSIL. 

The release process led to increased transparency. 
We dispelled knowledge asymmetries concerning performance limitations and the implementation status. 
Harmonization documents like operating instructions promoted internal communication quality, as they served as drivers of binding decisions both for the development and the demonstration setting. 
Although posing a burden to developers, extensive documentation enabled close monitoring of the safety concept's realization, allowing us to reveal potential deficits in advance and impede realizing design decisions based thereon that could have led to hazardous prototype behavior.
So, while preparing a release decision by disclosing the risk reduction actually achieved by implementation was the main goal when conceptualizing the release process, sensitizing the developers to safety-relevant considerations also had a risk-reducing effect during prototype development.

Project partners, who focused on function development or on system safety respectively, worked closely together to ensure the fulfillment of safety requirements. 
This close collaboration revealed, among other things, that developers of specific domains might have divergent understandings when considering ``safety'' in an implementation context. 
Thereby, the process highlighted the relevance of a debate about the definition of ``safety'' but also could show that a process tailored to nurture internal communication contributes to the harmonization of stakeholder-individual understandings.

Involving a certification agency added great value. 
The feedback on accompanied tests and the review of evidence we provided increased overall confidence in the safety measures.
Our system-wide safety analyses were found to align with established standards and processes to an adequate degree. 
Yet, although a release recommendation represents an indicator for appropriate measures taken to reduce risk, an external assessment should not be the sole reason for release. 
Release documents must still be thoroughly reviewed\footnote{The process execution in the project yielded around 700 pages long release documents per prototype for the public demonstration.} by the release committee to avoid potential confirmation bias.

We encountered a trade-off in defining release modules that are required for a release stage. 
A demand too great slows down development progress since more effort, which is required to release prototypes in an early phase, leads to delayed testing of new functions. 
If the demand is too low, the release process fails to serve its purpose.
Release documents will then provide insufficient evidence that adequate risk-reducing mechanisms have been implemented and tested. 
One approach that has proven helpful is to consult the system architecture to aid the determination of functional components required for the operation in question.
However, ways to establish traceability between release modules and a prototype's architecture are still subject to our research. 
In our future work, we intend to investigate following questions:
\begin{itemize}
	\item To what extent can formerly released components be reused in a related context?
	\item During operation, what triggers call an already-granted release on component or vehicle level into question?
	\item How do we, aided by architectural considerations, promote the tracing of triggers to affected release modules? 
	\item What are consequences of system modifications for the prototype operation formerly approved?
	\item How to prepare requirements and test cases to have advanced knowledge about necessary regression tests?
\end{itemize}


\section{CONCLUSION}
\label{sec:Conclusion}

While automated vehicle prototypes are already operating in urban environments on a daily basis, there is hardly any literature on a systematic release for prototypes. 
Based on our experience from the project \unicaragil, we aim to narrow this gap. 
Hence, we presented a release process for prototypes' demonstration that involves coordinated release process actor collaboration. 
The process follows a Safety-by-Design paradigm and targets challenges of prototype development and deployment in a research context.

In this paper, we argued the suitability of the process to reveal the achieved risk reduction.
According to gained knowledge, we deem coherent release documents, enriched by all actors and prepared adequately, as an appropriate basis for a prototype release decision. 
It became apparent that a close supervision of function development by safety engineers helps uncover deficits and design conflicts. 

The process execution has shown that a common understanding of ``safety'' among all those involved in development supports the release of prototypes.
While the release process poses a procedural framework to prepare the basis for a release and drive the development process in a safety-oriented manner, release parties also must be able to deduce when they can assess residual risk exposed in release documents as tolerable.
Thus, there is a need to lead a debate on the safety level during development (i.e., for the use of prototypes) which we want to initiate with this paper.
As part of the \autotech\ project, we will investigate the stated research needs and aim to advance the presented process.


\section*{ACKNOWLEDGMENT}

We thank Sonja Luther and Niklas Braun for proofreading, Torben Stolte for his contributions to the presented work, and Udo Steiniger for his valuable input in the course of a contracted review of the safety concept in the project.


\renewcommand*{\bibfont}{\footnotesize}

\printbibliography

@article{thorpe1988,
	author = {Thorpe, Chuck and Hebert, Martial and Kanade, Takeo and Shafer, Steven},
	title = {Vision and {Navigation} for the {Carnegie-Mellon} {Navlab}},
	journal = {IEEE Transactions on Pattern Analysis and Machine Intelligence},
	year = {1988},
	month = {May},
	volume = {10},
	number = {3},
	pages = {362 - 373},
}

@misc{AFGBV2022,
	title        = {{Autonome-Fahrzeuge-Genehmigungs-und-Betriebs-Verordnung} - {AFGBV}},
	howpublished = {BGBI. I p. 986},
	year = {2022},
	month = jun,
	note = {Verordnung zur Genehmigung und zum Betrieb von Kraftfahrzeugen mit autonomer Fahrfunktion in festgelegten Betriebsbereichen}
}

@article{dickmanns1987,
	title = {Autonomous {High} {Speed} {Road} {Vehicle} {Guidance} by {Computer} {Vision}},
	journal = {IFAC Proceedings Volumes},
	volume = {20},
	number = {5, Part 4},
	pages = {221-226},
	year = {1987},
	note = {10th Triennial IFAC Congress on Automatic Control - 1987 Volume IV, Munich, Germany},
	issn = {1474-6670},
	url = {https://www.sciencedirect.com/science/article/pii/S1474667017553203},
	author = {E.D. Dickmanns and A. Zapp}
}

@mis{DMV2023,
	author={{California State Transportation Agency -- Department of Motor Vehicles}},
	title = {Order of {Suspension} ({Cruise} {LLC})},
	note = {[Online]. Available: https://s3.documentcloud.org/documents/24080715/gm-cruise-order-of-suspension-driverless-testing.pdf},
	year ={2023}
}

@inproceedings{jatzkowski2021,
	title = {{Integration} of a {Vehicle} {Operating} {Mode} {Management} into {UNICARagil's} {Automotive} {Service-oriented} {Software} {Architecture}},
	booktitle = {30th Aachen Colloquium Sustainable Mobility},
	author = {Jatzkowski, Inga and Stolte, Torben and Graubohm, Robert and Maurer, Markus and Kampmann, Alexandru and Alrifaee, Bassam and Kowalewski, Stefan and Buchholz, Michael and Dietmayer, Klaus},
	date = {2021},
	doi = {10.24355/dbbs.084-202110271613-0},
	month = oct,
	pages = {595-614},
	address = {Aachen, Germany}
}

@misc{homann2002,
	title = {{Wirtschaft} und gesellschaftliche {Akzeptanz}: {Fahrerassistenzsysteme} auf dem {Prüfstand}},
	type = {Presentation},
	note = {Uni-DAS e.V. Workshop Fahrerassistenz, Walting, Germany},
	author = {Homann, Karl},
	year = {2002}
}

@inproceedings{strauss2023,
	title = {A {Pragmatic {Approach} {to} {Safe} {Operation} {for} {Driverless} {Shuttles} {During} {Development}}},
	author = {Strau{\ss}, Matthias and Pinke, Christopher},
	date = {2023},
	booktitle = {27th International Technical Conference on the Enhanced Safety of Vehicles}
}

@inproceedings{nothdurft2011,
	title = {Stadtpilot: {{First}} fully autonomous test drives in urban traffic},
	shorttitle = {Stadtpilot},
	booktitle = {14th {{International IEEE Conference}} on {{Intelligent Transportation Systems}} ({{ITSC}})},
	author = {Nothdurft, Tobias and Hecker, Peter and Ohl, Sebastian and Saust, Falko and Maurer, Markus and Reschka, Andreas and Bohmer, Jurgen Rudiger},
	year = {2011},
	month = oct,
	pages = {919--924},
	publisher = {{IEEE}},
	address = {{Washington, DC, USA}},
	doi = {10.1109/ITSC.2011.6082883},
	url = {http://ieeexplore.ieee.org/document/6082883/},
	urldate = {2023-12-14},
	isbn = {978-1-4577-2197-7}
}

@inproceedings{stolte2020,
	title = {Towards {{Safety Concepts}} for {{Automated Vehicles}} by the {{Example}} of the {{Project UNICARagil}}},
	author = {Stolte, Torben and Graubohm, Robert and Jatzkowski, Inga and Maurer, Markus and Ackermann, Stefan and Klamann, Bj{\"o}rn and Lippert, Moritz and Winner, Hermann},
	date = {2020},
	doi = {10.24355/dbbs.084-202011171557-0},
	booktitle = {29th Aachen Colloquium Sustainable Mobility},
	pages = {1561-1594}
}

@inproceedings{vankempen2023,
	author = {{van Kempen}, Raphael and Lampe, Bastian and Leuffen, Marc and Wirtz, Lena and Thomsen, Fabian and Bilkei-Gorzo, Gergely and Busch, Jean-Pierre and Feger, Ida and Geller, Christian and Kehl, Christian and Uszynski, Olaf and Wagner-Douglas, Lotte and Zanger, Lukas and Eckstein, Lutz and Klüner, David and Beerwerth, Julius and Alrifaee, Bassam and Kowalewski, Stefan and Konersmann, Marco and Dietmayer, Klaus},
	booktitle = {32nd Aachen Colloquium Sustainable Mobility},
	year = {2021},
	month = oct,
	pages = {1-49},
	title = {{AUTOtech.\emph{agil}}: {Architecture} and {Technologies} for {Orchestrating} {Automotive} {Agility}},
	doi = {10.18154/RWTH-2023-09783}
}

@inproceedings{graubohm2020b,
	title = {Towards {{Efficient Hazard Identification}} in the {{Concept Phase}} of {{Driverless Vehicle Development}}},
	booktitle = {{{IEEE Intelligent Vehicles Symposium}} ({{IV}})},
	author = {Graubohm, Robert and Stolte, Torben and Bagschik, Gerrit and Maurer, Markus},
	year = {2020},
	month = oct,
	pages = {1297--1304},
	publisher = {{IEEE}},
	address = {{Las Vegas, NV, USA}},
	doi = {10.1109/IV47402.2020.9304780},
	url = {https://ieeexplore.ieee.org/document/9304780/},
	urldate = {2024-02-22},
	isbn = {978-1-72816-673-5}
}

@inproceedings{jan2021,
	title = {Safety-configuration of {{Autonomous Bus}} in {{Pedestrian Zone}}:},
	shorttitle = {Safety-configuration of {{Autonomous Bus}} in {{Pedestrian Zone}}},
	booktitle = {Proceedings of the 7th {{International Conference}} on {{Vehicle Technology}} and {{Intelligent Transport Systems}}},
	author = {Jan, Qazi and Berns, Karsten},
	year = {2021},
	doi = {10.5220/0010526106980705},
	url = {https://www.scitepress.org/DigitalLibrary/Link.aspx?doi=10.5220/0010526106980705},
	urldate = {2023-12-14},
	isbn = {978-989-758-513-5}
}

@article{wansley2024,
	author={Wansley, Matthew},
	title={Regulating Driving Automation Safety},
	volume={73},
	journal={Emory Law Journal},
	year={2024}
}

@article{salem2024,
	author={Salem, Nayel Fabian and Kirschbaum, Thomas and Nolte, Marcus and Lalitsch-Schneider, Christian and Graubohm, Robert and Reich, Jan and Maurer, Markus},
	journal={IEEE Access}, 
	title={{Risk} {Management} {Core}---{Toward} an {Explicit} {Representation} of {Risk} in {Automated} {Driving}}, 
	year={2024},
	volume={12},
	number={},
	doi={10.1109/ACCESS.2024.3372860}
}

@article{broggi2013,
  title = {Extensive {{Tests}} of {{Autonomous Driving Technologies}}},
  author = {Broggi, Alberto and Buzzoni, Michele and Debattisti, Stefano and Grisleri, Paolo and Laghi, Maria Chiara and Medici, Paolo and Versari, Pietro},
  year = {2013},
  month = sep,
  journal = {IEEE Transactions on Intelligent Transportation Systems},
  volume = {14},
  number = {3},
  issn = {1524-9050, 1558-0016},
  doi = {10.1109/TITS.2013.2262331},
  url = {https://ieeexplore.ieee.org/document/6522193},
  urldate = {2023-12-14}
}

@book{zurawka2016,
	author = {Schäuffele, Jörg and Zurawka, Thomas},
	title = {Automotive Software Engineering},
	publisher = {{SAE International}},
	year = {2016},
	address = {{Warrendale, PA, USA}}
}

@incollection{salemTobepublished,
	title = {Safety and {Risk} -- {Why} {Their} {Definitions} {Matter}},
	booktitle = {Handbook {{Assisted}} and {{Automated Driving}}},
	author = {Salem, Nayel Fabian and Le Page, Sophie and Millar, Jason and Junietz, Philipp and Nolte, Marcus and Graubohm, Robert and Maurer, Markus},
	note = {unpublished},
	edition = {4},
	publisher = {{Springer Vieweg}},
	editor ={{Winner, Hermann, and Dietmayer, Klaus and Eckstein, Lutz and Jipp, Meike and Maurer, Markus and Stiller, Christoph}},
	address = {{Wiesbaden, Germany}}
}

@book{ISO_2018,
	title = {{Road} vehicles --- {Functional} safety},
	note = {{International} {Orgnization} for {Standardization} {Standard} 26262, 2018}
}

@book{vdi2006,
	title = {Design methodology for mechatronic systems},
	note={VDI Guideline 2206, {Verein} {Deutscher} {Ingenieure}, {Düsseldorf}, {Germany}, 2004}
}

@book{ISO_2022,
	title = {{{Road}} vehicles --- {{Safety}} of the intended functionality},
	note = {{International} {Orgnization} for {Standardization} {Standard} 21448, 2022}
}

@misc{maurer2023,
	title = {{Das {inhärente} {Risiko} {autonomer} {Straßenfahrzeuge}}},
	type = {Presentation},
	note = {{Braunschweig} {Mobility} {Days} -- {Autonom} und {Digital}, {Fachtagung} »{Autonomes} {Fahren} und {Stadtstruktur}«, Braunschweig, Germany},
	author = {Maurer, Markus},
	date = {2023-06-02},
	year = {2023},
	month = jun
}

@misc{fleischer2023,
	title = {{Safety} and {Acceptance} –- {A} {View} of {Two} {Mysteries}},
	type = {Presentation},
	note = {Oberseminar IfR Braunschweig, virtual},
	author = {Torsten Fleischer},
	year = {2023},
	month = sep
}

@techreport{maurer2018,
  title = {{Elektronische Fahrzeugsysteme -- {Jahresbericht}: Akademisches Jahr 2017/2018}},
  author = {Maurer, Markus},
  note ={Ed.: Gerrit Bagschik}
}

@techreport{maurer2021,
	title = {{Elektronische Fahrzeugsysteme -- {Jahresbericht}: Akademisches Jahr 2017/2018}},
	author = {Maurer, Markus},
	note ={Ed.: Inga Jatzkowski}
}

@techreport{bagschik2018,
	title = {aFAS -- {Automatisch} fahrerlos fahrendes {Absicherungsfahrzeug} für {Arbeitsstellen} auf {Bundesautobahnen}: {AP2} -- {Final Report}},
	author = {Bagschik, Gerrit and Steimle, Markus and Stolte, Torben and Maurer, Markus},
	year ={2019},
	doi={10.2314/GBV:1662494076}
}

@incollection{thorpe1998,
  title = {Automated {{Highways}} and the {{Free Agent Demonstration}}},
  booktitle = {Robotics {{Research}}},
  author = {Thorpe, Chuck and Jochem, Todd and Pomerleau, Dean},
  editor = {Shirai, Yoshiaki and Hirose, Shigeo},
  year = {1998},
  pages = {246--254},
  publisher = {{Springer London}},
  address = {{London}},
  doi = {10.1007/978-1-4471-1580-9_23},
  url = {http://link.springer.com/10.1007/978-1-4471-1580-9_23},
  urldate = {2023-12-14},
  isbn = {978-1-4471-1580-9}
}

@incollection{wachenfeld2016,
  title = {The {{Release}} of {{Autonomous Vehicles}}},
  booktitle = {{Autonomous} driving: {{Technical}}, legal and social aspects},
  author = {Wachenfeld, Walther and Winner, Hermann},
  editor = {Maurer, Markus and Gerdes, J. Christian and Lenz, Barbara and Winner, Hermann},
  year = {2016},
  pages = {425--449},
  publisher = {{Springer Berlin Heidelberg}},
  address = {{Berlin, Heidelberg}},
  doi = {10.1007/978-3-662-48847-8_21},
  url = {https://doi.org/10.1007/978-3-662-48847-8_21},
  isbn = {978-3-662-48847-8}
}

@phdthesis{ackermann2023,
	title={{Safe Halt} as {Fail-safe} {Concept} for {Automated} {Driving} {Systems}},
	author={Ackermann, Stefan},
	year={2023},
	school={Technische Universität Darmstadt},
	type = {Ph.D. dissertation},
	address = {Darmstadt, Germany}
}

@techreport{waymollc2021,
  title = {Waymo {{Safety Report}}},
  author = {{Waymo LLC}},
  year = {2021},
  month = feb,
  url = {https://downloads.ctfassets.net/e6t5diu0txbw/4mhzJxuCinbVNuyAKPPcOj/d1623d42ed7aaea46993c22ea7e50612/Waymo_Safety_Report_02-2021.pdf}
}

@article{ziegler2014,
  title = {Making {{Bertha Drive}}---{{An Autonomous Journey}} on a {{Historic Route}}},
  author = {Ziegler, Julius and Bender, Philipp and Schreiber, Markus and Lategahn, Henning and Strauss, Tobias and Stiller, Christoph and {Thao Dang} and Franke, Uwe and Appenrodt, Nils and Keller, Christoph G. and Kaus, Eberhard and Herrtwich, Ralf G. and Rabe, Clemens and Pfeiffer, David and Lindner, Frank and Stein, Fridtjof and Erbs, Friedrich and Enzweiler, Markus and Knoppel, Carsten and Hipp, Jochen and Haueis, Martin and Trepte, Maximilian and Brenk, Carsten and Tamke, Andreas and Ghanaat, Mohammad and Braun, Markus and Joos, Armin and Fritz, Hans and Mock, Horst and Hein, Martin and Zeeb, Eberhard},
  year = 2014,
  journal = {IEEE Intelligent Transportation Systems Magazine},
  volume = {6},
  number = {2},
  pages = {8--20},
  issn = {1939-1390},
  doi = {10.1109/MITS.2014.2306552},
  url = {http://ieeexplore.ieee.org/document/6803933/},
  urldate = {2023-12-14}
}

\end{document}